\documentclass[12pt, a4paper]{article}
\usepackage{graphicx}
\usepackage{amssymb}
\usepackage{amsmath}              
\usepackage{cases}         
\usepackage{cancel}              
\usepackage{bm}
\usepackage{color}
\usepackage{cite}
\usepackage{braket}
\usepackage[sort&compress,numbers, merge]{natbib} 
\usepackage{url} 
\usepackage{hyperref}

\newcommand{\1}{\mbox{1}\hspace{-0.25em}\mbox{l}}

\def\lrf#1#2{ \left(\frac{#1}{#2}\right)}
\def\lrfp#1#2#3{ \left(\frac{#1}{#2} \right)^{#3}}

\def\l{\left}
\def\r{\right}

\begin{document}

\begin{titlepage}

\begin{flushright}
TU-1063\\
IPMU 18-0080\\
MIT-CTP/5016
\end{flushright}
\vskip 2cm
\begin{center}

{\large
{\bf 
Heavy Gravitino from Dynamical Generation of Right-Handed Neutrino Mass Scale, and Gravitational Waves}
}

\vskip 2cm
Ryo Nagai$^{a}$,
Fuminobu Takahashi$^{a,b,c}$,
Norimi Yokozaki$^{a}$,

\vskip 0.5cm
{\it $^a$
Department of Physics, Tohoku University, Sendai, Miyagi
980-8578, Japan}\\[2pt]
{\it $^b$
Kavli Institute for the Physics and Mathematics of the Universe (WPI),\\
University of Tokyo, Kashiwa 277-8583, Japan}\\[2pt]
{\it $^c$ Department of Physics, Massachusetts Institute of Technology, \\
Cambridge, MA 02139 USA}

\date{\today}

\vskip 1.5cm

\begin{abstract} 
The absence of supersymmetric particles at the weak scale is a puzzle if supersymmetry solves
the gauge hierarchy problem. We show that, if the right-handed neutrino masses arise from 
hidden gaugino condensation via GUT-suppressed operators, successful thermal leptogenesis 
leads to the lower bound  on the gravitino mass, which may explain the little hierarchy problem. 
If domain walls associated with the gaugino condensation are formed after inflation, they
annihilate when $H \sim m_{3/2}$.
We address the possibility that observable gravitational wave signals might emerge from 
the domain wall annihilation.
\end{abstract}

\end{center}
\end{titlepage}

\section{Introduction}
 
 Supersymmetric extensions of the Standard Model (SUSY SM) has been one of the leading candidates for physics beyond the SM.
 It not only stabilizes the weak scale against radiative corrections,
but also leads to successful gauge coupling unification at $M_{\rm GUT} \sim 2 \times 10^{16}$\,GeV, hinting at the presence of a grand unified theory (GUT). However, no SUSY particles have been found at the LHC so far, pushing up the SUSY breaking scale to a few TeV or even heavier. 

The discovery of the SM-like Higgs boson with a mass of 125\,GeV \cite{Aad:2012tfa,Chatrchyan:2012xdj} is also consistent with such a 
high-scale SUSY breaking scenario. For instance, if all the sfermion masses are of order the gravitino mass heavier than about 10\,TeV~\cite{ArkaniHamed:2004fb,Giudice:2004tc,ArkaniHamed:2004yi,Wells:2004di, ArkaniHamed:2005yv,Giudice:2011cg,Hall:2011jd,Ibe:2011aa,ArkaniHamed:2012gw} , one can explain the SM-like Higgs boson mass thanks to the large stop radiative corrections~\cite{Okada:1990vk, Ellis:1990nz, Haber:1990aw, Okada:1990gg, Ellis:1991zd}. The high-scale SUSY scenario  greatly ameliorates the SUSY flavor and CP problems~\cite{Gabbiani:1996hi, Ellis:1982tk,Buchmuller:1982ye,delAguila:1983dfr,Polchinski:1983zd,Franco:1983xm} as well as the gravitino/moduli problems~\cite{Khlopov:1984pf,Ellis:1984eq,Kawasaki:2008qe, Coughlan:1983ci,Banks:1993en,deCarlos:1993wie,Endo:2006zj,Nakamura:2006uc}.

One of the central issues in the high-scale SUSY scenario is the little hierarchy problem. Namely, the question why the SUSY breaking scale is much higher than the weak scale, if the SUSY is to solve the gauge hierarchy problem. In this paper, we show that thermal leptogenesis may play a key role in solving the little hierarchy problem. 

Suppose that there is a hidden SUSY Yang-Mills (YM) sector that is coupled to right-handed neutrinos  through nonrenormalizable operators.  In the hidden SUSY YM sector, the fermionic partner of the hidden gauge boson, hidden gaugino, confines at low energy and forms
a condensate. Then, the right-handed neutrinos acquire their masses through the interactions with the gaugino condensate, and so 
their mass scale is related to the dynamical scale of the gaugino condensation.  The dynamical scale cannot be arbitrarily high, however,
 and its upper bound is related to the gravitino mass if we require the vanishingly small cosmological constant in the present vacuum. As we shall see, the gravitino mass is bounded below for a given right-handed neutrino mass scale. Interestingly,  if we take the effective cut-off scale around the GUT scale, the typical mass scale of the right-handed neutrinos required for successful thermal leptogenesis~\cite{Fukugita:1986hr}  leads the heavy gravitino mass, $m_{3/2} \gtrsim 10$\,TeV.  Thus, the little hierarchy may be a result of successful thermal leptogenesis.

In our setup, the hidden strong dynamics not only generates the mass of right-handed neutrino but also predict interesting phenomena in the early Universe.  To be concrete, let us consider a hidden SU(N) SUSY YM sector, where a U(1)$_R$ symmetry is explicitly broken down to $Z_{2N {\bf R}}$ symmetry.  At the dynamical scale, the discrete symmetry is further spontaneously broken to $Z_{2{\bf R}}$ by the gaugino condensation, leading to 
formation of the domain walls. If the gaugino condensate is formed after the inflation, the domain-wall network follows the scaling law. 
If the $Z_{2N {\bf R}}$ symmetry is also explicitly broken, it creates an energy
bias between different vacua.  When the pressure on the domain walls due to the energy bias becomes comparable to the
their tension,  the domain walls start to annihilate, and the gaugino condensate decays into the right-handed neutrinos. The minimal explicit $Z_{2N {\bf R}}$ breaking comes from the constant term in the superpotential, which determines the gravitino mass. 
If this is the main source of the energy bias, the domain walls decay at $H\simeq m_{3/2}$~\cite{Takahashi:2008mu}. In the course of violent annihilation processes, a large amount of gravitational waves can be emitted. The frequency and abundance of the gravitational waves we observe today are respectively determined by the gravitino mass and the dynamical scale. We will address a possibility to detect the gravitational waves by the forthcoming gravitational wave experiments such as Advanced LIGO~\cite{Abbott:2017xzg}, Einstein Telescope~\cite{Hild:2008ng}, and an experiment using Bose-Einstein condensation (BEC)~\cite{Sabin:2015mha}.

Lastly let us mention related works in the past~\cite{Babu:2015xba,Bjorkeroth:2016qsk}. 
In Ref.~\cite{Babu:2015xba}, they proposed a scenario where 
both U(1)$_{B-L}$ and SUSY are dynamically broken at the same time, which relates the right-handed neutrino mass 
to the gravitino mass, and then derived a lower-bound on the gravitino mass, $m_{3/2} \gtrsim \mathcal{O}(1000)$\,TeV, 
assuming non-thermal leptogenesis.
In Ref.~\cite{Bjorkeroth:2016qsk},  they studied resonant leptogenesis~\cite{Covi:1996fm,Pilaftsis:1997dr,Buchmuller:1997yu,Pilaftsis:2003gt,Pilaftsis:2005rv,Anisimov:2005hr,Garny:2011hg} in a context of the right-handed sneutrino 
inflation~\cite{Nakayama:2016gvg}, and found that the gravitino mass must be larger than $\mathcal{O}(100)$\,TeV.

The rest of this paper is organized as follows. In Sec. \ref{sec:The Model}, we give a model where the mass of the right-handed neutrino is
generated by a hidden gaugino condensate. Then we show that
the lower-bound on the gravitino mass is fixed by the requirement of the successful thermal leptogenesis.
In Sec.~\ref{sec:Cosmological implication}, we discuss cosmological implications including predictions for the gravitational wave signals.
Finally, Sec.~\ref{sec:Summary} is devoted to conclusions.  

\section{Right-handed neutrino mass and hidden gaugino condensate}
\label{sec:The Model}
The crucial ingredient of our model is that the heavy right-handed neutrinos acquire their mass from 
a spontaneous breaking of the U(1)$_R$ symmetry. 
In the seesaw mechanism~\cite{Yanagida:1979as,GellMann:1980vs,Minkowski:1977sc}, the right-handed 
neutrino mass term is usually allowed by symmetry, and so, it can be naturally of 
order $10^{14-15}$\,GeV.
Instead here we assume that the right-handed neutrino mass is 
forbidden by the U(1)$_R$ symmetry, and it is generated by the spontaneous breaking of the U(1)$_R$
symmetry such as gaugino condensation. 

To be explicit, let us consider a hidden SU(N) SUSY Yang-Mills sector. 
The gauge interactions become strong at a dynamical scale $\Lambda$,
and the hidden gaugino $\lambda^A_H$ forms a condensate,
\begin{align}
\braket{\lambda^A_H\lambda^A_H}=-32\pi^2\Lambda^3e^{2\pi ik/N},
\label{eq:GC}
\end{align}
where $k=1,\cdots N$ labels the $N$ distinct vacua~\cite{Shifman:1987ia}. 
Below the dynamical scale $\Lambda$, the effective superpotential of the gaugino condensation is given by
\begin{align}
W_{GC}
=
N\Lambda^3e^{2\pi ik/N}.
\label{WGC}
\end{align}
Now let us assume that the hidden gauge sector is coupled
to the right-handed neutrinos via GUT-suppressed operators,
\begin{align}
W_{N_R}
= \frac{1}{M_{\ast}^2}
N\Lambda^3e^{2\pi ik/N} \bar{N}_R \bar{N}_R
\label{WNN}
\end{align}
where $\bar{N}_R$ is a chiral superfield containing a right-handed neutrino, $M_{\ast}$ denotes the effective cut-off
scale for the interaction between the hidden sector and the right-handed neutrino. We assume that $M_*$ is of order the GUT scale, 
$M_* \sim$\,$10^{15}$\,-\,$10^{16}$\,GeV. Here and in what follows we suppress flavor indices for simplicity.
Therefore, the right-handed neutrino obtains its Majorana mass from the coupling to the hidden gaugino condensate,
\begin{align}
M_{N_R}
&= 2 \frac{N\Lambda^3}{M_*^2},\nonumber\\
&\simeq  2 \times 10^{10} {\rm\,GeV}  \lrf{ N \Lambda^3}{(10^{14}{\rm\,GeV})^3}
\lrfp{10^{16}{\rm\,GeV}}{M_*}{2},
\label{eq:Majorana}
\end{align}
up to a phase factor. One example of U(1)$_R$ charge assignment is given at the end of this section.

The scale of the gaugino condensation is bounded above, and its upper bound is related to the gravitino mass.
To see this, we note that the superpotential  contains a constant term, $W_0$,  to cancel the positive contribution of the SUSY breaking
to the cosmological constant,\footnote{We have assumed $\left<K\right> \ll M_P^2$, where $K$ is a Kahler potential.}
\begin{align}
W_0
= N \Lambda^3 e^{2\pi ik/N} + {w}_0
\equiv m_{3/2} M_{\rm{Pl}}^2 e^{i \delta},
\label{W0}
\end{align}
where $w_0$ is the constant term, $m_{3/2}$ the gravitino mass, and $M_{\rm{Pl}}=2.4\times 10^{18}$\,GeV  the reduced Plank mass.
Here the gravitino mass is defined as a real parameter,  and
we have introduced a complex phase factor, $e^{i \delta}$.
 Barring cancellation among various contributions to the constant term, we obtain
\begin{align}
R \equiv \frac{N \Lambda^3}{m_{3/2} M_{\rm{Pl}}^2} \lesssim 1
\label{ineq}
\end{align}
Combining Eqs.~(\ref{WGC}), (\ref{eq:Majorana}),  (\ref{W0}), and (\ref{ineq}), we arrive at
\begin{align}
m_{3/2}
\gtrsim 
8\,\mbox{TeV}
\l(\frac{M_{\ast}}{10^{16}\,\mbox{GeV}}\r)^2
\l(\frac{M_{N_R}}{10^{9}\,\mbox{GeV}}\r).
\end{align}
Successful thermal leptogenesis requires $M_{N_R} \gtrsim 4 \times 10^{8}$\,GeV, assuming 
thermal initial abundance~\cite{Buchmuller:2002jk}. Thus, successful thermal leptogenesis implies that
the gravitino mass $m_{3/2}$ should be heavier than several TeV for the GUT-scale cut-off.

A few comments are in order. First, the assumption about the initial
abundance is likely satisfied in a scenario to be discussed in the
next section, where a large number of right-handed neutrinos (as well as gravitational waves) 
is produced by the domain wall annihilation. Secondly, if the right-handed neutrino masses are slightly degenerate,
the resultant baryon asymmetry can be enhanced by the resonant leptogenesis~\cite{Covi:1996fm,Pilaftsis:1997dr,Buchmuller:1997yu,Pilaftsis:2003gt,Pilaftsis:2005rv,Anisimov:2005hr,Garny:2011hg}. Therefore, the lower bound on the gravitino mass can be
relaxed in this case. Thirdly, 
the gravitino is known to cause the gravitino problem~\cite{Khlopov:1984pf,Ellis:1984eq,Kawasaki:2008qe}: the gravitino is thermally 
produced in the early Universe, and it later decays into the visible particles during the big bang nucleosynthesis, modifying the light
element abundances in contradiction
with the observation. If the gravitino is the lightest SUSY particle (LSP), its abundance may exceed the dark matter abundance. 
The gravitino problem is relaxed significantly if the gravitino mass is heavier than ${\cal O}(10)$\,TeV, as it decays before the
big bang nucleosynthesis, or if there is another dark sector (e.g. dark photon and dark photino) 
which contains the LSP with a mass much lighter than the gravitino mass. One can also avoid the overproduction of 
the LSP produced from the gravitino decay by introducing a small R-parity violating operator~\cite{Takayama:2000uz,Buchmuller:2007ui}.

Before closing this section, let us give a concrete R-charge assignment of the matter fields
which forbids the right-handed neutrino mass.
Here we adopt a $Z_{6{\bf R}}$ symmetry with the $R$-charge of $R(\bar{N}_R)=3$.
See Table~\ref{tab:model} for the R-charges of the other SSM fields. 
 The right-handed neutrino can couple to the SSM sector in the $Z_{6{\bf R}}$ symmetric manner 
 as~{\footnote{Here we have suppressed $\mathcal{O}$(1) coefficients of these terms.}}
\begin{align}
W
=
Q\bar{U}H_u
+
Q\bar{D}H_d
+
L\bar{E}H_d
+
L\bar{N}_R H_u,
\end{align} 
while the bare Majorana mass is forbidden by the symmetry. Then, the Majorana mass is induced by the gaugino condensation,
which spontaneously breaks  $Z_{6{\bf R}}$ down to $Z_{2{\bf R}}$, where $W_{GC}$ has an effective $R$-charge $2$.
Note that the $\mu$-term is also forbidden
by the symmetry.\footnote{Another discrete R-symmetry, $Z_{4{\bf R}}$, also forbids the $\mu$-term (see e.g.~\cite{Lee:2010gv}).
However, it allows the bare Majorana mass term.
}
 In fact, one can generate the $\mu$-term similarly by introducing
non-renormalizable coupling to the hidden gaugino condensation,
$W = W_{GC}^2 H_u H_d/M_*^5$. For $\Lambda\sim 10^{14}$\,GeV and $M_* \sim 10^{16}$\,GeV,
one can obtain $\mu \sim {\cal O}(1)$\,TeV.

\begin{table}[t!]
 \begin{center}
\caption{The matter contents and the charge assignment}
\label{tab:model}
\vspace{5pt}
\begin{tabular}{ccccccccc}
\hline
 & $Q$ & $\bar{U}$ & $\bar{D}$ & $L$ & $\bar{E}$ & $\bar{N}_R$ & $H_u$ & $H_d$ \\
\hline
\hline
$SU(3)_C$ & ${\bf{3}}$ & $\bar{{\bf{3}}}$ & $\bar{{\bf{3}}}$ & ${\bf{1}}$ & ${\bf{1}}$ & ${\bf{1}}$ & ${\bf{1}}$ & ${\bf{1}}$ \\
$SU(2)_L$ & ${\bf{2}}$ & ${\bf{1}}$ & ${\bf{1}}$ & ${\bf{2}}$ & ${\bf{1}}$ & ${\bf{1}}$ & ${\bf{2}}$ & ${\bf{2}}$ \\
$U(1)_Y$ & $1/6$ & $-2/3$ & $1/3$ & $-1/2$ & $1$ & $0$ & $1/2$ & $-1/2$ \\
\hline
$Z_{6{\bf R}}$ & $3$ & $3$ & $3$ & $3$ & $3$ & $3$ & $2$ & $2$ \\
\hline
\end{tabular}
 \end{center}
\end{table}

\section{Cosmological implications}
\label{sec:Cosmological implication}
Next we discuss the cosmological aspects of our model. 
In the hidden gauge sector, the continuous $R$ symmetry is  anomalous, and it is explicitly 
broken to a discrete $Z_{2N}$ subgroup by non-perturbative effects \cite{Witten:1982df}. Furthermore, 
at the dynamical scale $\Lambda$, 
the discrete symmetry is spontaneously broken down to $Z_{2}$ symmetry due to the gaugino condensation,
leading to the domain wall 
formation \cite{Zeldovich:1974uw,Kibble:1976sj,Vilenkin:1981zs,Dvali:1996xe,Matsuda:1998ms}.

Let us suppose that domain walls are formed after inflation. This is the case if~\footnote{
If the Hubble parameter during inflation is lower than the dynamical scale,
we consider a case in which the hidden sector is in equilibrium with the SSM sector after the reheating 
or it is heated by the inflaton decay up to a temperature above the dynamical scale. 
If domain walls are formed before inflation, they are diluted away by the subsequent inflation,
and there is no cosmological impact on the observable Universe.
}
\begin{align}
\max({H_{\rm{inf}},T_{\rm{max}}})\gtrsim\Lambda
\label{eq:cond2}
\end{align}
is satisfied, where  $H_{\rm{inf}}$ denotes the Hubble parameter during the inflation. 
$T_{\rm{max}}$ denotes the highest temperature of the background plasma 
given by \cite{Giudice:2000ex,Asaka:1999xd}
\begin{align}
T_{\rm{max}}\simeq (T^2_{\rm{reh}}H_{\rm{inf}}M_{\rm{Pl}})^{1/4},
\end{align}
where $T_{\rm{reh}}$ is the reheating temperature. 
The domain walls are known to follow the scaling law, and their energy density decreases more slowly than radiation or matter.
Therefore, if the domain walls were completely stable, 
they would dominate the Universe at the end of the day and generate intolerably large spatial inhomogeneities and anisotropies~\cite{Vilenkin:1981zs}.
To avoid the cosmological catastrophe, one needs to introduce an energy bias between the different vacua, which makes
the domain walls unstable and disappear at a later time. The domain wall annihilation is so violent that a large amount of gravitational waves
can be produced.  As we shall see below, the domain walls annihilate when the Hubble parameter
becomes comparable to the gravitino mass, which determines the peak frequency of the gravitational waves.

The domain walls are formed when either the Hubble parameter or cosmic temperature becomes equal to $\Lambda$. 
The tension of the domain wall between the adjacent vacua, $\sigma_{\rm DW}$ is given by~\cite{Dvali:1996xe}
\begin{align}
\sigma_{\rm{DW}}
&= 
2N \Lambda^3|e^{2\pi i/N}-1|
\leq
4N\Lambda^3.
\end{align}
Hereafter we take $\sigma_{\rm{DW}}=4N\Lambda^3$ as a reference value.
After formation, the network of the domain walls is known to quickly follow the scaling law.
The energy density of the domain wall $\rho_{\rm{DW}}$ in the scaling regime is estimated as
\begin{align}
\rho_{\rm{DW}}(t)
&=
\mathcal{A} \frac{\sigma_{\rm{DW}}}{t},
\end{align}
where $\mathcal{A}$ is an $\mathcal{O}(1)$ parameter \cite{Hiramatsu:2012sc,Hiramatsu:2013qaa},
and $t$ is the cosmic time. Here and in what follows we assume the radiation dominated Universe, $H = 1/2t$,
unless otherwise stated.
The energy density of the domain walls decreases more slowly than radiation, and so, it comes to dominate
the Universe when
\begin{align}
t_{\rm dom} = \frac{3 M_{\rm{Pl}^2}}{4 \mathcal{A}\, \sigma_{\rm DW}}. 
\end{align}
To avoid the cosmological catastrophe, domain walls must decay before $t_{\rm dom}$.

The constant term in the superpotential provides the energy bias between different vacua.\footnote{
If the constant term in the superpotential is generated by another hidden gaugino condensate, 
 the condensation scale should be larger than the Hubble parameter during inflation so that the 
domain walls are diluted away by inflation~\cite{Dine:2010eb,Harigaya:2015yla}.
}
The relevant part of the scalar potential in supergravity reads
\begin{align}
V_{\rm bias} &= - \frac{3}{M_{\rm{Pl}^2} }N \Lambda^3 w_0^* e^{2\pi i/N} + h.c.,\nonumber \\
&=-6N m_{3/2} \Lambda^3 (\cos \theta_k - R)
\end{align}
where we have defined $\theta_k \equiv \delta - 2\pi k/N$.
The energy bias between the $k$-th and $(k+1)$-th vacua is then given by
\begin{align}
\epsilon_{\rm bias}^{k,k+1} = 6N m_{3/2} \Lambda^3 (\cos \theta_k - \cos \theta_{k+1}).
\end{align}
In the following analysis, we take a typical energy bias $\epsilon_{\rm bias}$ as $6N m_{3/2} \Lambda^3$.
The domain walls annihilate when their energy density becomes comparable to the bias energy density at $t = t_{\rm{ann}}$.
This is given by
\begin{align}
t_{\rm{ann}}
=
C_{d}\mathcal{A}\frac{\sigma_{\rm{DW}}}{\epsilon_{\rm{bias}}},
\end{align} 
where $C_d$ is a coefficient of $\mathcal{O}(1)$ \cite{Kawasaki:2014sqa} and 
$t_{\rm ann} \sim m_{3/2}^{-1}$. 

After the domain walls annihilate, the gaugino condensate mainly decays into a pair of right-handed neutrinos 
through the interaction (\ref{WNN}). The decay rate is roughly estimated as
\begin{align}
\Gamma_{\rm GC} \sim \frac{M_{N_R}^2}{8 \pi \Lambda} 
\end{align}
If this is larger than the Hubble parameter at the domain wall annihilation, the gaugino condensation instantly
decays into the right-handed neutrinos. Otherwise, it would take some time for the gaugino condensation to decay,
and it may even dominate the Universe and produces an entropy. 
To see this more explicitly, let us evaluate the amount of the entropy dilution factor.
The gaugino condensation after the domain wall decay is non-relativistic, and so, its energy density decreases
like matter. Therefore, it would dominate the Universe when the Hubble parameter becomes equal to
\begin{align}
H_{\rm GC, dom} \simeq \frac{4 \mathcal{A}^2 \sigma_{\rm DW}^2}{9 H_{\rm ann} M_{\rm Pl}^4},
\end{align}
where $H_{\rm ann} = 1/2 t_{\rm{ann}}$. Therefore, if $\Gamma_{\rm GC} > H_{\rm GC, dom}$,
there is no entropy production. On the other hand, if $\Gamma_{\rm GC} < H_{\rm GC, dom}$,
the gaugino condensation dominates the Universe, and its decay increases the entropy by a factor
$\Delta \equiv (H_{\rm GC, dom}/\Gamma_{\rm GC})^{1/2}$.

The collapses of the domain walls generate the gravitational wave \cite{Gleiser:1998na}.
The peak frequency of the gravitational wave  is determined by
the Hubble parameter $H_{\rm{ann}}$ which is the typical curvature radius of the domain wall:
\begin{align}
f_{\rm peak}
\simeq
H_{\rm{ann}},
\end{align}
and the energy density of the gravitational wave 
is estimated as
\begin{align}
\rho_{\rm{GW}}
\sim
G\mathcal{A}^2\sigma^2_{\rm{DW}},
\label{eq:EGW}
\end{align}
where $G=(8\pi M^2_{\rm{Pl}})^{-1}$ is the newton constant. 
The amplitudes of the gravitational wave for a frequency $f$ can be parameterized using the following dimensionless quantity: 
\begin{align}
\Omega_{\rm{GW}} (f)
\equiv
\frac{1}{\rho_c}\frac{d\rho_{\rm{GW}}}{d\log f},
\end{align}
where $\rho_c$ the critical energy density. Then, the peak amplitude is estimated as~\cite{Hiramatsu:2013qaa} 
\begin{eqnarray}
\Omega_{\rm{GW}} (f_{\rm peak}) = \frac{8\pi \tilde{\epsilon}_{\rm GW} G^2 \mathcal{A}^2 \sigma_{\rm DW}^2 }{3 H_{\rm ann}^2} ,
\end{eqnarray}
where $\tilde{\epsilon}_{\rm GW}$ around unity is an efficiency parameter for the gravitational radiation.
To estimate the property of the gravitational wave we observe today, we have to take into account the red-shift effect. The amount of the effect is given by $a(t_{\rm{ann}})/a_0$ where $a_0$ and $a(t_{\rm{ann}})$ denote the scale factors at present and at the formation of the gravitational waves, respectively. 
If the gravitational waves are created after the reheating, the red-shift parameter is given as
\begin{align}
\frac{a(t_{\rm{ann}})}{a_0}
\simeq
6\times 10^{-14}\l(\frac{g_\ast(t_{\rm{ann}})}{200}\r)^{-\frac{1}{3}}\l(\frac{T_{\rm{ann}}}{\mbox{GeV}}\r)^{-1},
\end{align}
where $T_{\rm{ann}}$ and $g_\ast(t_{\rm{ann}})$ are the temperature and the relativistic degrees of freedom at $H=H_{\rm{ann}}$. Thus, the peak frequency and the energy fraction observed today are estimated as
\begin{align}  
f_0 = f_{\rm peak}|_{\rm{present}}
\simeq
3.7 \times 10^4\,\mbox{Hz}\,(C_d\mathcal{A})^{-\frac{1}{2}}\l(\frac{g_\ast(t_{\rm{ann}})}{200}\r)^{-\frac{1}{12}}
\l(\frac{m_{3/2}}{100{\rm\,TeV}}\r)^{\frac{1}{2}},
\label{eq:f-afterRH}
\end{align}
and
\begin{align}  
\Omega_{\rm{GW}}(f_0) |_{\rm{present}}
\simeq
5 \times10^{-8}\, \tilde{\epsilon}_{\rm{GW}}C^2_d\mathcal{A}^4 N
\l(\frac{g_\ast(t_{\rm{ann}})}{200}\r)^{-\frac{1}{12}} \lrfp{R}{0.1}{2}.
\label{eq:Omega-afterRH}
\end{align}
On the other hand, if there is a period of the gaugino condensation domination after the domain wall annihilation,
the frequency and the density parameter of the gravitational waves are reduced by a factor of $\Delta^{1/3}$
and $\Delta^{4/3}$, respectively. In the following analysis, we take $\mathcal{A}=0.9$, $C_d=5$, $\tilde{\epsilon}_{\rm{GW}}=1.5$, and $N=5$.

\begin{figure}[!t]
\begin{center}
\includegraphics[width=79mm]{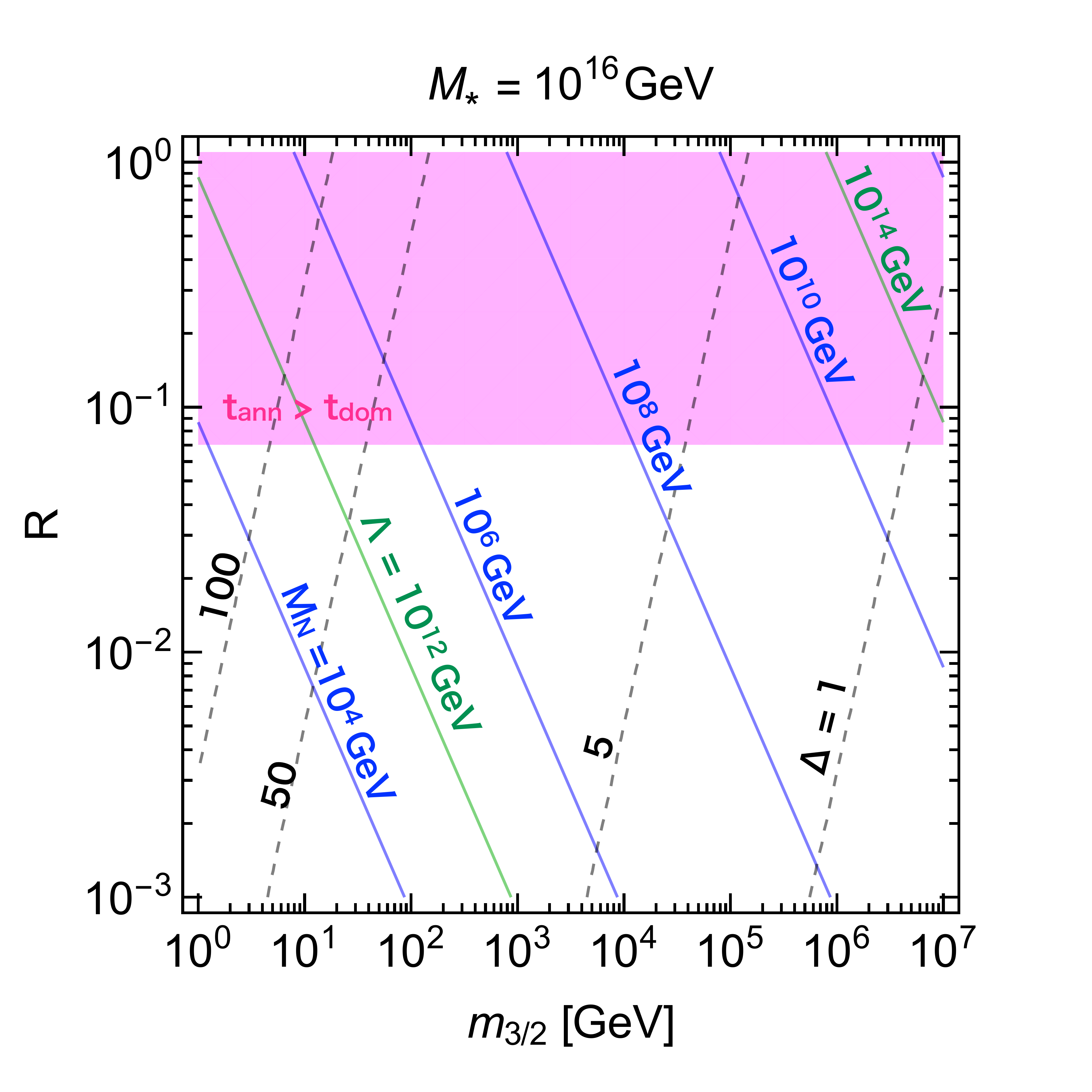}
\includegraphics[width=79mm]{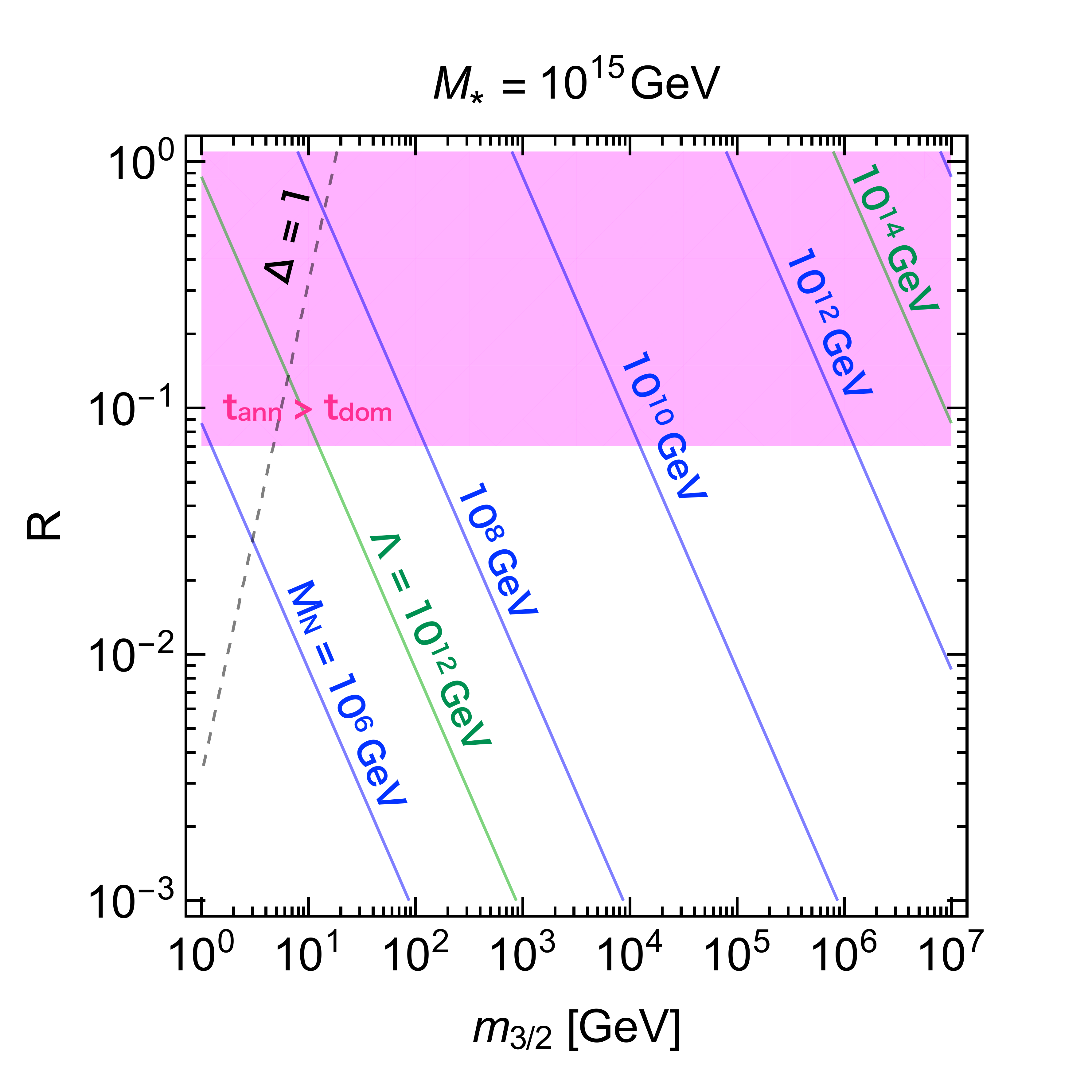}
\end{center}
\caption{Contours of the dynamical scale $\Lambda$ (green), right-handed neutrino mass $M_{N_R}$ (blue) and 
the entropy dilution factor $\Delta$ (gray dashed)
on the plane of $(m_{3/2}, R)$ for $M_* = 10^{16}$\,GeV (left) and $10^{15}$\,GeV (right). 
}
\label{fig:region}
\end{figure}

\begin{figure}[!t]
\begin{center}
\includegraphics[width=150mm]{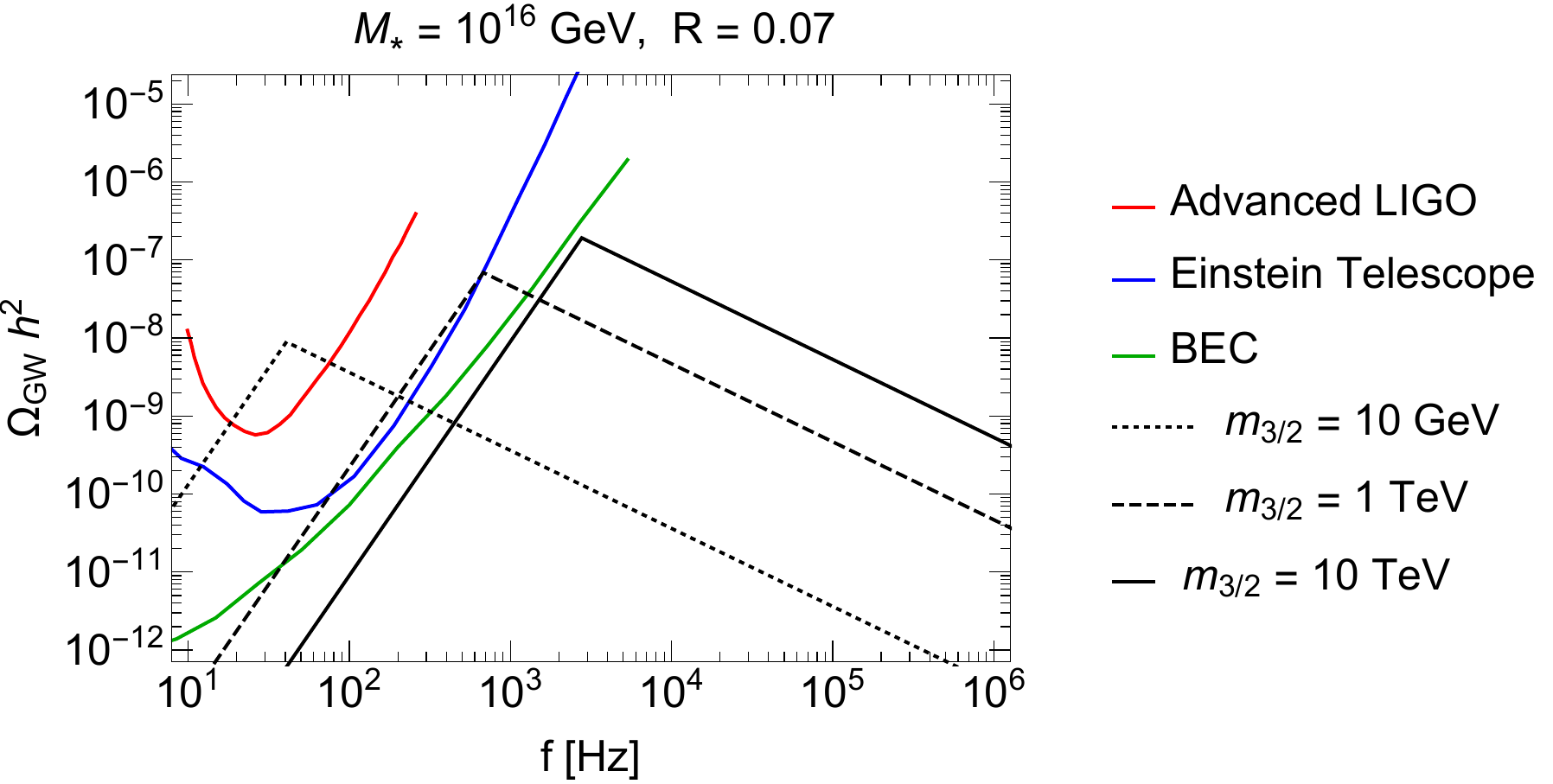}
\includegraphics[width=150mm]{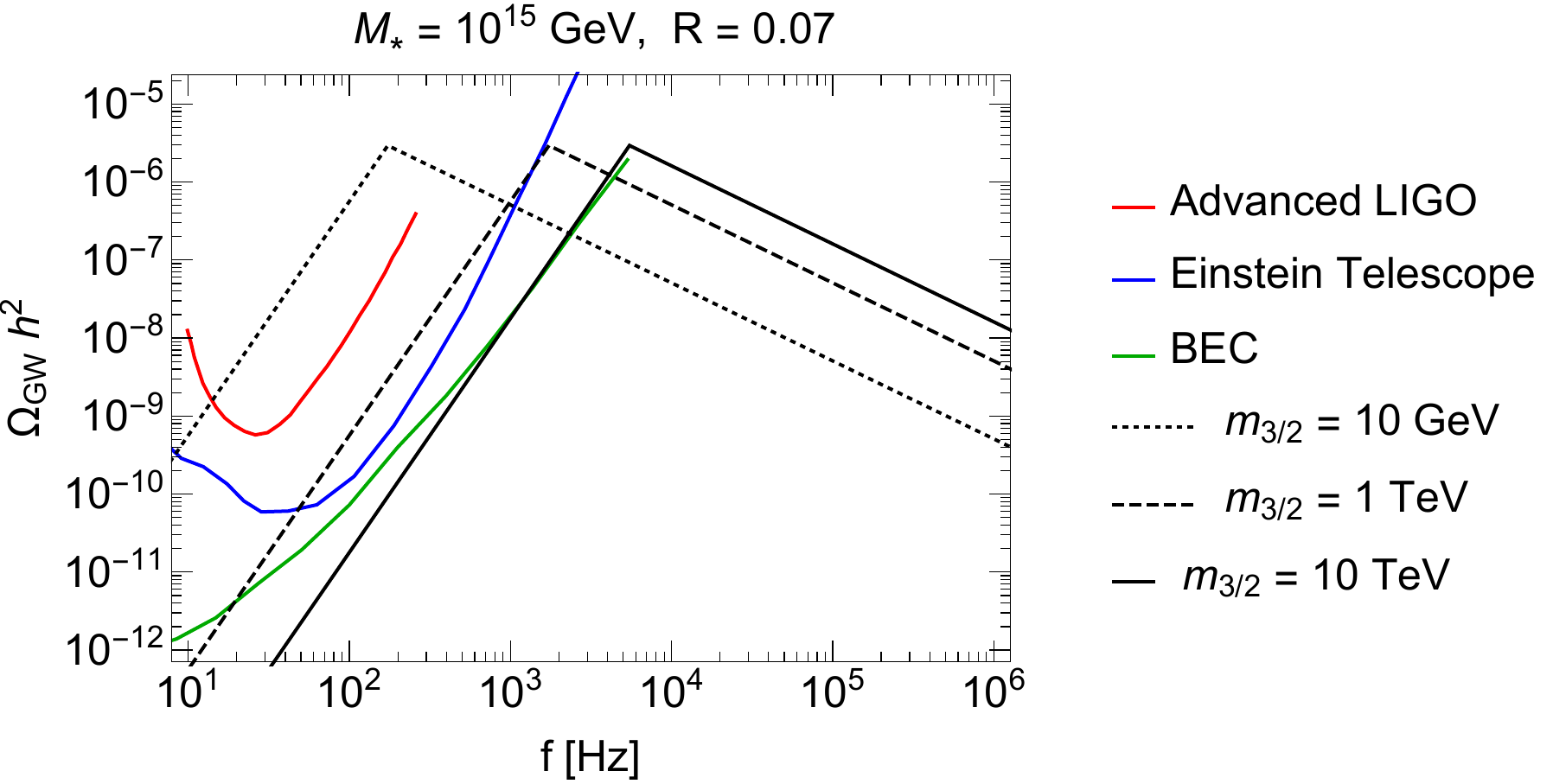}
\end{center}
\caption{Estimated gravitational wave spectra for $m_{3/2} = 10$\,TeV (solid), $1$\,TeV (dashed), and $10$\,GeV (dotted).
We also show the projected sensitivity reach of Advanced LIGO (red), Einstein Telescope (blue), and the experiment using BEC (green). The lines for $m_{3/2} = 10$\,GeV are shown for the comparison purpose. 
}
\label{fig:GW}
\end{figure}

In Fig.~\ref{fig:region}, We show the contours of the dynamical scale $\Lambda$, right-handed neutrino mass $M_{N_R}$, and dilution factor $\Delta$. The magenta regions are excluded due to $t_{\rm dom} < t_{\rm ann}$, namely the domain wall dominates the Universe before its annihilation.  In the left(right) panel, we take $M_{\ast}=10^{16}$\,GeV ($M_{\ast}=10^{15}$\,GeV). Including the cosmological constraints, $M_{N_R} \gtrsim 10^9$\,GeV leads to the lower-bound on $m_{3/2}$ as $10^5$\,GeV ($10^3$\,GeV) for $M_*=10^{16}$\,GeV ($M_*=10^{15}$\,GeV).
One can see that the entropy dilution factor is at most a factor of $5$ or so in the viable parameter space where successful
thermal leptogenesis takes place.

Figure \ref{fig:GW} shows the predicted gravitational wave spectra and the projected sensitivity reach of
the current and future detectors (Advanced LIGO, Einstein Telescope (ET), and the experiment using BEC). 
On the top (bottom) panel, we take $M_{\ast}=10^{16}$\,GeV ($M_{\ast}=10^{15}$\,GeV).
The peak frequency and energy fraction are determined by Eq.~(\ref{eq:f-afterRH}) and (\ref{eq:Omega-afterRH}),
together with the dilution factor $\Delta$.
We assume that the scale dependence of the energy fraction is $\Omega_{\rm{GW}}\propto f^3$ ($\Omega_{\rm{GW}}\propto f^{-1}$) for $f<f_0$ ($f>f_0$) according to Ref.~\cite{Hiramatsu:2012sc}. 
The sensitivity lines of $\Omega_{\rm{GW}}$ are taken from Ref.~\cite{Abbott:2017xzg} for Advanced LIGO, Ref.~\cite{Hild:2008ng} for ET, and Ref.~\cite{Sabin:2015mha} for BEC. The case of $m_{3/2} = 10$\,GeV is shown for the comparison purpose.\footnote{The gravitino LSP and
the little hierarchy problem were discussed in Ref.~\cite{Takahashi:2011as}.}
One can see that for $M_*=10^{15}$\,{\rm GeV}, the gravitational wave can be detected by ET and BEC
 in the viable region consistent with successful thermal leptogenesis.

\section{Summary}
\label{sec:Summary}
We have shown that, if the right-handed neutrino mass arises from the hidden gaugino condensation
through the GUT-suppressed interactions,  successful thermal leptogenesis scenario requires the gravitino mass 
heavier than several TeV. Thus, the little hierarchy problem may be explained by thermal leptogenesis.
Interestingly, if the gaugino condensation appears after inflation, domain walls
may come close to dominating the Universe before they decay. As a result, a significant amount of gravitational
waves is produced by the violent annihilation processes, and the peak frequency of the gravitational waves
is determined by the gravitino mass. We have shown that the predicted gravitational waves are out of the 
Advanced LIGO, but could be within reach of Einstein Telescope and the experiment using BEC for 
the gravitino mass of ${\cal O}(1)$\,TeV.

\section*{Acknowledgments}
F.T. thanks the hospitality of MIT Center for Theoretical Physics where part of this work was done.
This work is supported by JSPS KAKENHI Grant Numbers JP15H05889 (F.T. and N.Y.),
JP15K21733 (F.T. and N.Y.), JP17H02878 (F.T.), JP17H05396 (N.Y.), JP17H02875 (F.T. and N.Y.),
and 16H06490 (R.N.),  Leading Young Researcher Overseas Visit Program at Tohoku University (F.T.),
and by World Premier International Research Center Initiative (WPI Initiative),
MEXT, Japan (F.T.).

\newpage
{\small

\begin{thebibliography}{68}%
\makeatletter
\providecommand \@ifxundefined [1]{%
 \@ifx{#1\undefined}
}%
\providecommand \@ifnum [1]{%
 \ifnum #1\expandafter \@firstoftwo
 \else \expandafter \@secondoftwo
 \fi
}%
\providecommand \@ifx [1]{%
 \ifx #1\expandafter \@firstoftwo
 \else \expandafter \@secondoftwo
 \fi
}%
\providecommand \natexlab [1]{#1}%
\providecommand \enquote  [1]{``#1''}%
\providecommand \bibnamefont  [1]{#1}%
\providecommand \bibfnamefont [1]{#1}%
\providecommand \citenamefont [1]{#1}%
\providecommand \href@noop [0]{\@secondoftwo}%
\providecommand \href [0]{\begingroup \@sanitize@url \@href}%
\providecommand \@href[1]{\@@startlink{#1}\@@href}%
\providecommand \@@href[1]{\endgroup#1\@@endlink}%
\providecommand \@sanitize@url [0]{\catcode `\\12\catcode `\$12\catcode
  `\&12\catcode `\#12\catcode `\^12\catcode `\_12\catcode `\%12\relax}%
\providecommand \@@startlink[1]{}%
\providecommand \@@endlink[0]{}%
\providecommand \url  [0]{\begingroup\@sanitize@url \@url }%
\providecommand \@url [1]{\endgroup\@href {#1}{\urlprefix }}%
\providecommand \urlprefix  [0]{URL }%
\providecommand \Eprint [0]{\href }%
\providecommand \doibase [0]{http://dx.doi.org/}%
\providecommand \selectlanguage [0]{\@gobble}%
\providecommand \bibinfo  [0]{\@secondoftwo}%
\providecommand \bibfield  [0]{\@secondoftwo}%
\providecommand \translation [1]{[#1]}%
\providecommand \BibitemOpen [0]{}%
\providecommand \bibitemStop [0]{}%
\providecommand \bibitemNoStop [0]{.\EOS\space}%
\providecommand \EOS [0]{\spacefactor3000\relax}%
\providecommand \BibitemShut  [1]{\csname bibitem#1\endcsname}%
\let\auto@bib@innerbib\@empty
\bibitem [{\citenamefont {Aad}\ \emph {et~al.}(2012)\citenamefont {Aad} \emph
  {et~al.}}]{Aad:2012tfa}%
  \BibitemOpen
  \bibfield  {author} {\bibinfo {author} {\bibfnamefont {G.}~\bibnamefont
  {Aad}} \emph {et~al.} (\bibinfo {collaboration} {ATLAS}),\ }\href {\doibase
  10.1016/j.physletb.2012.08.020} {\bibfield  {journal} {\bibinfo  {journal}
  {Phys. Lett.}\ }\textbf {\bibinfo {volume} {B716}},\ \bibinfo {pages} {1}
  (\bibinfo {year} {2012})},\ \Eprint
  {http://arxiv.org/abs/1207.7214}{arXiv:1207.7214 [hep-ex]}\BibitemShut
  {NoStop}%
\bibitem [{\citenamefont {Chatrchyan}\ \emph {et~al.}(2012)\citenamefont
  {Chatrchyan} \emph {et~al.}}]{Chatrchyan:2012xdj}%
  \BibitemOpen
  \bibfield  {author} {\bibinfo {author} {\bibfnamefont {S.}~\bibnamefont
  {Chatrchyan}} \emph {et~al.} (\bibinfo {collaboration} {CMS}),\ }\href
  {\doibase 10.1016/j.physletb.2012.08.021} {\bibfield  {journal} {\bibinfo
  {journal} {Phys. Lett.}\ }\textbf {\bibinfo {volume} {B716}},\ \bibinfo
  {pages} {30} (\bibinfo {year} {2012})},\ \Eprint
  {http://arxiv.org/abs/1207.7235}{arXiv:1207.7235 [hep-ex]}\BibitemShut
  {NoStop}%
\bibitem [{\citenamefont {Arkani-Hamed}\ and\ \citenamefont
  {Dimopoulos}(2005)}]{ArkaniHamed:2004fb}%
  \BibitemOpen
  \bibfield  {author} {\bibinfo {author} {\bibfnamefont {N.}~\bibnamefont
  {Arkani-Hamed}}\ and\ \bibinfo {author} {\bibfnamefont {S.}~\bibnamefont
  {Dimopoulos}},\ }\href {\doibase 10.1088/1126-6708/2005/06/073} {\bibfield
  {journal} {\bibinfo  {journal} {JHEP}\ }\textbf {\bibinfo {volume} {06}},\
  \bibinfo {pages} {073} (\bibinfo {year} {2005})},\ \Eprint
  {http://arxiv.org/abs/hep-th/0405159}{arXiv:hep-th/0405159
  [hep-th]}\BibitemShut {NoStop}%
\bibitem [{\citenamefont {Giudice}\ and\ \citenamefont
  {Romanino}(2004)}]{Giudice:2004tc}%
  \BibitemOpen
  \bibfield  {author} {\bibinfo {author} {\bibfnamefont {G.~F.}\ \bibnamefont
  {Giudice}}\ and\ \bibinfo {author} {\bibfnamefont {A.}~\bibnamefont
  {Romanino}},\ }\href {\doibase 10.1016/j.nuclphysb.2004.11.048,
  10.1016/j.nuclphysb.2004.08.001} {\bibfield  {journal} {\bibinfo  {journal}
  {Nucl. Phys.}\ }\textbf {\bibinfo {volume} {B699}},\ \bibinfo {pages} {65}
  (\bibinfo {year} {2004})},\ \bibinfo {note} {[Erratum: Nucl.
  Phys.B706,487(2005)]},\ \Eprint
  {http://arxiv.org/abs/hep-ph/0406088}{arXiv:hep-ph/0406088
  [hep-ph]}\BibitemShut {NoStop}%
\bibitem [{\citenamefont {Arkani-Hamed}\ \emph
  {et~al.}(2005{\natexlab{a}})\citenamefont {Arkani-Hamed}, \citenamefont
  {Dimopoulos}, \citenamefont {Giudice},\ and\ \citenamefont
  {Romanino}}]{ArkaniHamed:2004yi}%
  \BibitemOpen
  \bibfield  {author} {\bibinfo {author} {\bibfnamefont {N.}~\bibnamefont
  {Arkani-Hamed}}, \bibinfo {author} {\bibfnamefont {S.}~\bibnamefont
  {Dimopoulos}}, \bibinfo {author} {\bibfnamefont {G.~F.}\ \bibnamefont
  {Giudice}}, \ and\ \bibinfo {author} {\bibfnamefont {A.}~\bibnamefont
  {Romanino}},\ }\href {\doibase 10.1016/j.nuclphysb.2004.12.026} {\bibfield
  {journal} {\bibinfo  {journal} {Nucl. Phys.}\ }\textbf {\bibinfo {volume}
  {B709}},\ \bibinfo {pages} {3} (\bibinfo {year} {2005}{\natexlab{a}})},\
  \Eprint {http://arxiv.org/abs/hep-ph/0409232}{arXiv:hep-ph/0409232
  [hep-ph]}\BibitemShut {NoStop}%
\bibitem [{\citenamefont {Wells}(2005)}]{Wells:2004di}%
  \BibitemOpen
  \bibfield  {author} {\bibinfo {author} {\bibfnamefont {J.~D.}\ \bibnamefont
  {Wells}},\ }\href {\doibase 10.1103/PhysRevD.71.015013} {\bibfield  {journal}
  {\bibinfo  {journal} {Phys. Rev.}\ }\textbf {\bibinfo {volume} {D71}},\
  \bibinfo {pages} {015013} (\bibinfo {year} {2005})},\ \Eprint
  {http://arxiv.org/abs/hep-ph/0411041}{arXiv:hep-ph/0411041
  [hep-ph]}\BibitemShut {NoStop}%
\bibitem [{\citenamefont {Arkani-Hamed}\ \emph
  {et~al.}(2005{\natexlab{b}})\citenamefont {Arkani-Hamed}, \citenamefont
  {Dimopoulos},\ and\ \citenamefont {Kachru}}]{ArkaniHamed:2005yv}%
  \BibitemOpen
  \bibfield  {author} {\bibinfo {author} {\bibfnamefont {N.}~\bibnamefont
  {Arkani-Hamed}}, \bibinfo {author} {\bibfnamefont {S.}~\bibnamefont
  {Dimopoulos}}, \ and\ \bibinfo {author} {\bibfnamefont {S.}~\bibnamefont
  {Kachru}},\ }\href@noop {} {\  (\bibinfo {year} {2005}{\natexlab{b}})},\
  \Eprint {http://arxiv.org/abs/hep-th/0501082}{arXiv:hep-th/0501082
  [hep-th]}\BibitemShut {NoStop}%
\bibitem [{\citenamefont {Giudice}\ and\ \citenamefont
  {Strumia}(2012)}]{Giudice:2011cg}%
  \BibitemOpen
  \bibfield  {author} {\bibinfo {author} {\bibfnamefont {G.~F.}\ \bibnamefont
  {Giudice}}\ and\ \bibinfo {author} {\bibfnamefont {A.}~\bibnamefont
  {Strumia}},\ }\href {\doibase 10.1016/j.nuclphysb.2012.01.001} {\bibfield
  {journal} {\bibinfo  {journal} {Nucl. Phys.}\ }\textbf {\bibinfo {volume}
  {B858}},\ \bibinfo {pages} {63} (\bibinfo {year} {2012})},\ \Eprint
  {http://arxiv.org/abs/1108.6077}{arXiv:1108.6077 [hep-ph]}\BibitemShut
  {NoStop}%
\bibitem [{\citenamefont {Hall}\ and\ \citenamefont
  {Nomura}(2012)}]{Hall:2011jd}%
  \BibitemOpen
  \bibfield  {author} {\bibinfo {author} {\bibfnamefont {L.~J.}\ \bibnamefont
  {Hall}}\ and\ \bibinfo {author} {\bibfnamefont {Y.}~\bibnamefont {Nomura}},\
  }\href {\doibase 10.1007/JHEP01(2012)082} {\bibfield  {journal} {\bibinfo
  {journal} {JHEP}\ }\textbf {\bibinfo {volume} {01}},\ \bibinfo {pages} {082}
  (\bibinfo {year} {2012})},\ \Eprint
  {http://arxiv.org/abs/1111.4519}{arXiv:1111.4519 [hep-ph]}\BibitemShut
  {NoStop}%
\bibitem [{\citenamefont {Ibe}\ and\ \citenamefont
  {Yanagida}(2012)}]{Ibe:2011aa}%
  \BibitemOpen
  \bibfield  {author} {\bibinfo {author} {\bibfnamefont {M.}~\bibnamefont
  {Ibe}}\ and\ \bibinfo {author} {\bibfnamefont {T.~T.}\ \bibnamefont
  {Yanagida}},\ }\href {\doibase 10.1016/j.physletb.2012.02.034} {\bibfield
  {journal} {\bibinfo  {journal} {Phys. Lett.}\ }\textbf {\bibinfo {volume}
  {B709}},\ \bibinfo {pages} {374} (\bibinfo {year} {2012})},\ \Eprint
  {http://arxiv.org/abs/1112.2462}{arXiv:1112.2462 [hep-ph]}\BibitemShut
  {NoStop}%
\bibitem [{\citenamefont {Arkani-Hamed}\ \emph {et~al.}(2012)\citenamefont
  {Arkani-Hamed}, \citenamefont {Gupta}, \citenamefont {Kaplan}, \citenamefont
  {Weiner},\ and\ \citenamefont {Zorawski}}]{ArkaniHamed:2012gw}%
  \BibitemOpen
  \bibfield  {author} {\bibinfo {author} {\bibfnamefont {N.}~\bibnamefont
  {Arkani-Hamed}}, \bibinfo {author} {\bibfnamefont {A.}~\bibnamefont {Gupta}},
  \bibinfo {author} {\bibfnamefont {D.~E.}\ \bibnamefont {Kaplan}}, \bibinfo
  {author} {\bibfnamefont {N.}~\bibnamefont {Weiner}}, \ and\ \bibinfo {author}
  {\bibfnamefont {T.}~\bibnamefont {Zorawski}},\ }\href@noop {} {\  (\bibinfo
  {year} {2012})},\ \Eprint {http://arxiv.org/abs/1212.6971}{arXiv:1212.6971
  [hep-ph]}\BibitemShut {NoStop}%
\bibitem [{\citenamefont {Okada}\ \emph
  {et~al.}(1991{\natexlab{a}})\citenamefont {Okada}, \citenamefont
  {Yamaguchi},\ and\ \citenamefont {Yanagida}}]{Okada:1990vk}%
  \BibitemOpen
  \bibfield  {author} {\bibinfo {author} {\bibfnamefont {Y.}~\bibnamefont
  {Okada}}, \bibinfo {author} {\bibfnamefont {M.}~\bibnamefont {Yamaguchi}}, \
  and\ \bibinfo {author} {\bibfnamefont {T.}~\bibnamefont {Yanagida}},\ }\href
  {\doibase 10.1143/ptp/85.1.1} {\bibfield  {journal} {\bibinfo  {journal}
  {Prog. Theor. Phys.}\ }\textbf {\bibinfo {volume} {85}},\ \bibinfo {pages}
  {1} (\bibinfo {year} {1991}{\natexlab{a}})}\BibitemShut {NoStop}%
\bibitem [{\citenamefont {Ellis}\ \emph
  {et~al.}(1991{\natexlab{a}})\citenamefont {Ellis}, \citenamefont {Ridolfi},\
  and\ \citenamefont {Zwirner}}]{Ellis:1990nz}%
  \BibitemOpen
  \bibfield  {author} {\bibinfo {author} {\bibfnamefont {J.~R.}\ \bibnamefont
  {Ellis}}, \bibinfo {author} {\bibfnamefont {G.}~\bibnamefont {Ridolfi}}, \
  and\ \bibinfo {author} {\bibfnamefont {F.}~\bibnamefont {Zwirner}},\ }\href
  {\doibase 10.1016/0370-2693(91)90863-L} {\bibfield  {journal} {\bibinfo
  {journal} {Phys. Lett.}\ }\textbf {\bibinfo {volume} {B257}},\ \bibinfo
  {pages} {83} (\bibinfo {year} {1991}{\natexlab{a}})}\BibitemShut {NoStop}%
\bibitem [{\citenamefont {Haber}\ and\ \citenamefont
  {Hempfling}(1991)}]{Haber:1990aw}%
  \BibitemOpen
  \bibfield  {author} {\bibinfo {author} {\bibfnamefont {H.~E.}\ \bibnamefont
  {Haber}}\ and\ \bibinfo {author} {\bibfnamefont {R.}~\bibnamefont
  {Hempfling}},\ }\href {\doibase 10.1103/PhysRevLett.66.1815} {\bibfield
  {journal} {\bibinfo  {journal} {Phys. Rev. Lett.}\ }\textbf {\bibinfo
  {volume} {66}},\ \bibinfo {pages} {1815} (\bibinfo {year}
  {1991})}\BibitemShut {NoStop}%
\bibitem [{\citenamefont {Okada}\ \emph
  {et~al.}(1991{\natexlab{b}})\citenamefont {Okada}, \citenamefont
  {Yamaguchi},\ and\ \citenamefont {Yanagida}}]{Okada:1990gg}%
  \BibitemOpen
  \bibfield  {author} {\bibinfo {author} {\bibfnamefont {Y.}~\bibnamefont
  {Okada}}, \bibinfo {author} {\bibfnamefont {M.}~\bibnamefont {Yamaguchi}}, \
  and\ \bibinfo {author} {\bibfnamefont {T.}~\bibnamefont {Yanagida}},\ }\href
  {\doibase 10.1016/0370-2693(91)90642-4} {\bibfield  {journal} {\bibinfo
  {journal} {Phys. Lett.}\ }\textbf {\bibinfo {volume} {B262}},\ \bibinfo
  {pages} {54} (\bibinfo {year} {1991}{\natexlab{b}})}\BibitemShut {NoStop}%
\bibitem [{\citenamefont {Ellis}\ \emph
  {et~al.}(1991{\natexlab{b}})\citenamefont {Ellis}, \citenamefont {Ridolfi},\
  and\ \citenamefont {Zwirner}}]{Ellis:1991zd}%
  \BibitemOpen
  \bibfield  {author} {\bibinfo {author} {\bibfnamefont {J.~R.}\ \bibnamefont
  {Ellis}}, \bibinfo {author} {\bibfnamefont {G.}~\bibnamefont {Ridolfi}}, \
  and\ \bibinfo {author} {\bibfnamefont {F.}~\bibnamefont {Zwirner}},\ }\href
  {\doibase 10.1016/0370-2693(91)90626-2} {\bibfield  {journal} {\bibinfo
  {journal} {Phys. Lett.}\ }\textbf {\bibinfo {volume} {B262}},\ \bibinfo
  {pages} {477} (\bibinfo {year} {1991}{\natexlab{b}})}\BibitemShut {NoStop}%
\bibitem [{\citenamefont {Gabbiani}\ \emph {et~al.}(1996)\citenamefont
  {Gabbiani}, \citenamefont {Gabrielli}, \citenamefont {Masiero},\ and\
  \citenamefont {Silvestrini}}]{Gabbiani:1996hi}%
  \BibitemOpen
  \bibfield  {author} {\bibinfo {author} {\bibfnamefont {F.}~\bibnamefont
  {Gabbiani}}, \bibinfo {author} {\bibfnamefont {E.}~\bibnamefont {Gabrielli}},
  \bibinfo {author} {\bibfnamefont {A.}~\bibnamefont {Masiero}}, \ and\
  \bibinfo {author} {\bibfnamefont {L.}~\bibnamefont {Silvestrini}},\ }\href
  {\doibase 10.1016/0550-3213(96)00390-2} {\bibfield  {journal} {\bibinfo
  {journal} {Nucl. Phys.}\ }\textbf {\bibinfo {volume} {B477}},\ \bibinfo
  {pages} {321} (\bibinfo {year} {1996})},\ \Eprint
  {http://arxiv.org/abs/hep-ph/9604387}{arXiv:hep-ph/9604387
  [hep-ph]}\BibitemShut {NoStop}%
\bibitem [{\citenamefont {Ellis}\ \emph {et~al.}(1982)\citenamefont {Ellis},
  \citenamefont {Ferrara},\ and\ \citenamefont {Nanopoulos}}]{Ellis:1982tk}%
  \BibitemOpen
  \bibfield  {author} {\bibinfo {author} {\bibfnamefont {J.~R.}\ \bibnamefont
  {Ellis}}, \bibinfo {author} {\bibfnamefont {S.}~\bibnamefont {Ferrara}}, \
  and\ \bibinfo {author} {\bibfnamefont {D.~V.}\ \bibnamefont {Nanopoulos}},\
  }\href {\doibase 10.1016/0370-2693(82)90484-1} {\bibfield  {journal}
  {\bibinfo  {journal} {Phys. Lett.}\ }\textbf {\bibinfo {volume} {114B}},\
  \bibinfo {pages} {231} (\bibinfo {year} {1982})}\BibitemShut {NoStop}%
\bibitem [{\citenamefont {Buchmuller}\ and\ \citenamefont
  {Wyler}(1983)}]{Buchmuller:1982ye}%
  \BibitemOpen
  \bibfield  {author} {\bibinfo {author} {\bibfnamefont {W.}~\bibnamefont
  {Buchmuller}}\ and\ \bibinfo {author} {\bibfnamefont {D.}~\bibnamefont
  {Wyler}},\ }\bibfield  {booktitle} {\emph {\bibinfo {booktitle} {{1982 DESY
  Workshop: Electroweak Interactions at High Energies Hamburg, Germany, Sept
  28-30, 1982}}},\ }\href {\doibase 10.1016/0370-2693(83)91378-3} {\bibfield
  {journal} {\bibinfo  {journal} {Phys. Lett.}\ }\textbf {\bibinfo {volume}
  {121B}},\ \bibinfo {pages} {321} (\bibinfo {year} {1983})},\ \bibinfo {note}
  {[,277(1982)]}\BibitemShut {NoStop}%
\bibitem [{\citenamefont {del Aguila}\ \emph {et~al.}(1983)\citenamefont {del
  Aguila}, \citenamefont {Gavela}, \citenamefont {Grifols},\ and\ \citenamefont
  {Mendez}}]{delAguila:1983dfr}%
  \BibitemOpen
  \bibfield  {author} {\bibinfo {author} {\bibfnamefont {F.}~\bibnamefont {del
  Aguila}}, \bibinfo {author} {\bibfnamefont {M.~B.}\ \bibnamefont {Gavela}},
  \bibinfo {author} {\bibfnamefont {J.~A.}\ \bibnamefont {Grifols}}, \ and\
  \bibinfo {author} {\bibfnamefont {A.}~\bibnamefont {Mendez}},\ }\href
  {\doibase 10.1016/0370-2693(83)90018-7} {\bibfield  {journal} {\bibinfo
  {journal} {Phys. Lett.}\ }\textbf {\bibinfo {volume} {126B}},\ \bibinfo
  {pages} {71} (\bibinfo {year} {1983})},\ \bibinfo {note} {[Erratum: Phys.
  Lett.129B,473(1983)]}\BibitemShut {NoStop}%
\bibitem [{\citenamefont {Polchinski}\ and\ \citenamefont
  {Wise}(1983)}]{Polchinski:1983zd}%
  \BibitemOpen
  \bibfield  {author} {\bibinfo {author} {\bibfnamefont {J.}~\bibnamefont
  {Polchinski}}\ and\ \bibinfo {author} {\bibfnamefont {M.~B.}\ \bibnamefont
  {Wise}},\ }\href {\doibase 10.1016/0370-2693(83)91310-2} {\bibfield
  {journal} {\bibinfo  {journal} {Phys. Lett.}\ }\textbf {\bibinfo {volume}
  {125B}},\ \bibinfo {pages} {393} (\bibinfo {year} {1983})}\BibitemShut
  {NoStop}%
\bibitem [{\citenamefont {Franco}\ and\ \citenamefont
  {Mangano}(1984)}]{Franco:1983xm}%
  \BibitemOpen
  \bibfield  {author} {\bibinfo {author} {\bibfnamefont {E.}~\bibnamefont
  {Franco}}\ and\ \bibinfo {author} {\bibfnamefont {M.~L.}\ \bibnamefont
  {Mangano}},\ }\href {\doibase 10.1016/0370-2693(84)90312-5} {\bibfield
  {journal} {\bibinfo  {journal} {Phys. Lett.}\ }\textbf {\bibinfo {volume}
  {135B}},\ \bibinfo {pages} {445} (\bibinfo {year} {1984})}\BibitemShut
  {NoStop}%
\bibitem [{\citenamefont {Khlopov}\ and\ \citenamefont
  {Linde}(1984)}]{Khlopov:1984pf}%
  \BibitemOpen
  \bibfield  {author} {\bibinfo {author} {\bibfnamefont {M.~{\relax Yu}.}\
  \bibnamefont {Khlopov}}\ and\ \bibinfo {author} {\bibfnamefont {A.~D.}\
  \bibnamefont {Linde}},\ }\href {\doibase 10.1016/0370-2693(84)91656-3}
  {\bibfield  {journal} {\bibinfo  {journal} {Phys. Lett.}\ }\textbf {\bibinfo
  {volume} {138B}},\ \bibinfo {pages} {265} (\bibinfo {year}
  {1984})}\BibitemShut {NoStop}%
\bibitem [{\citenamefont {Ellis}\ \emph {et~al.}(1984)\citenamefont {Ellis},
  \citenamefont {Kim},\ and\ \citenamefont {Nanopoulos}}]{Ellis:1984eq}%
  \BibitemOpen
  \bibfield  {author} {\bibinfo {author} {\bibfnamefont {J.~R.}\ \bibnamefont
  {Ellis}}, \bibinfo {author} {\bibfnamefont {J.~E.}\ \bibnamefont {Kim}}, \
  and\ \bibinfo {author} {\bibfnamefont {D.~V.}\ \bibnamefont {Nanopoulos}},\
  }\href {\doibase 10.1016/0370-2693(84)90334-4} {\bibfield  {journal}
  {\bibinfo  {journal} {Phys. Lett.}\ }\textbf {\bibinfo {volume} {145B}},\
  \bibinfo {pages} {181} (\bibinfo {year} {1984})}\BibitemShut {NoStop}%
\bibitem [{\citenamefont {Kawasaki}\ \emph {et~al.}(2008)\citenamefont
  {Kawasaki}, \citenamefont {Kohri}, \citenamefont {Moroi},\ and\ \citenamefont
  {Yotsuyanagi}}]{Kawasaki:2008qe}%
  \BibitemOpen
  \bibfield  {author} {\bibinfo {author} {\bibfnamefont {M.}~\bibnamefont
  {Kawasaki}}, \bibinfo {author} {\bibfnamefont {K.}~\bibnamefont {Kohri}},
  \bibinfo {author} {\bibfnamefont {T.}~\bibnamefont {Moroi}}, \ and\ \bibinfo
  {author} {\bibfnamefont {A.}~\bibnamefont {Yotsuyanagi}},\ }\href {\doibase
  10.1103/PhysRevD.78.065011} {\bibfield  {journal} {\bibinfo  {journal} {Phys.
  Rev.}\ }\textbf {\bibinfo {volume} {D78}},\ \bibinfo {pages} {065011}
  (\bibinfo {year} {2008})},\ \Eprint
  {http://arxiv.org/abs/0804.3745}{arXiv:0804.3745 [hep-ph]}\BibitemShut
  {NoStop}%
\bibitem [{\citenamefont {Coughlan}\ \emph {et~al.}(1983)\citenamefont
  {Coughlan}, \citenamefont {Fischler}, \citenamefont {Kolb}, \citenamefont
  {Raby},\ and\ \citenamefont {Ross}}]{Coughlan:1983ci}%
  \BibitemOpen
  \bibfield  {author} {\bibinfo {author} {\bibfnamefont {G.~D.}\ \bibnamefont
  {Coughlan}}, \bibinfo {author} {\bibfnamefont {W.}~\bibnamefont {Fischler}},
  \bibinfo {author} {\bibfnamefont {E.~W.}\ \bibnamefont {Kolb}}, \bibinfo
  {author} {\bibfnamefont {S.}~\bibnamefont {Raby}}, \ and\ \bibinfo {author}
  {\bibfnamefont {G.~G.}\ \bibnamefont {Ross}},\ }\href {\doibase
  10.1016/0370-2693(83)91091-2} {\bibfield  {journal} {\bibinfo  {journal}
  {Phys. Lett.}\ }\textbf {\bibinfo {volume} {131B}},\ \bibinfo {pages} {59}
  (\bibinfo {year} {1983})}\BibitemShut {NoStop}%
\bibitem [{\citenamefont {Banks}\ \emph {et~al.}(1994)\citenamefont {Banks},
  \citenamefont {Kaplan},\ and\ \citenamefont {Nelson}}]{Banks:1993en}%
  \BibitemOpen
  \bibfield  {author} {\bibinfo {author} {\bibfnamefont {T.}~\bibnamefont
  {Banks}}, \bibinfo {author} {\bibfnamefont {D.~B.}\ \bibnamefont {Kaplan}}, \
  and\ \bibinfo {author} {\bibfnamefont {A.~E.}\ \bibnamefont {Nelson}},\
  }\href {\doibase 10.1103/PhysRevD.49.779} {\bibfield  {journal} {\bibinfo
  {journal} {Phys. Rev.}\ }\textbf {\bibinfo {volume} {D49}},\ \bibinfo {pages}
  {779} (\bibinfo {year} {1994})},\ \Eprint
  {http://arxiv.org/abs/hep-ph/9308292}{arXiv:hep-ph/9308292
  [hep-ph]}\BibitemShut {NoStop}%
\bibitem [{\citenamefont {de~Carlos}\ \emph {et~al.}(1993)\citenamefont
  {de~Carlos}, \citenamefont {Casas}, \citenamefont {Quevedo},\ and\
  \citenamefont {Roulet}}]{deCarlos:1993wie}%
  \BibitemOpen
  \bibfield  {author} {\bibinfo {author} {\bibfnamefont {B.}~\bibnamefont
  {de~Carlos}}, \bibinfo {author} {\bibfnamefont {J.~A.}\ \bibnamefont
  {Casas}}, \bibinfo {author} {\bibfnamefont {F.}~\bibnamefont {Quevedo}}, \
  and\ \bibinfo {author} {\bibfnamefont {E.}~\bibnamefont {Roulet}},\ }\href
  {\doibase 10.1016/0370-2693(93)91538-X} {\bibfield  {journal} {\bibinfo
  {journal} {Phys. Lett.}\ }\textbf {\bibinfo {volume} {B318}},\ \bibinfo
  {pages} {447} (\bibinfo {year} {1993})},\ \Eprint
  {http://arxiv.org/abs/hep-ph/9308325}{arXiv:hep-ph/9308325
  [hep-ph]}\BibitemShut {NoStop}%
\bibitem [{\citenamefont {Endo}\ \emph {et~al.}(2006)\citenamefont {Endo},
  \citenamefont {Hamaguchi},\ and\ \citenamefont {Takahashi}}]{Endo:2006zj}%
  \BibitemOpen
  \bibfield  {author} {\bibinfo {author} {\bibfnamefont {M.}~\bibnamefont
  {Endo}}, \bibinfo {author} {\bibfnamefont {K.}~\bibnamefont {Hamaguchi}}, \
  and\ \bibinfo {author} {\bibfnamefont {F.}~\bibnamefont {Takahashi}},\ }\href
  {\doibase 10.1103/PhysRevLett.96.211301} {\bibfield  {journal} {\bibinfo
  {journal} {Phys. Rev. Lett.}\ }\textbf {\bibinfo {volume} {96}},\ \bibinfo
  {pages} {211301} (\bibinfo {year} {2006})},\ \Eprint
  {http://arxiv.org/abs/hep-ph/0602061}{arXiv:hep-ph/0602061
  [hep-ph]}\BibitemShut {NoStop}%
\bibitem [{\citenamefont {Nakamura}\ and\ \citenamefont
  {Yamaguchi}(2006)}]{Nakamura:2006uc}%
  \BibitemOpen
  \bibfield  {author} {\bibinfo {author} {\bibfnamefont {S.}~\bibnamefont
  {Nakamura}}\ and\ \bibinfo {author} {\bibfnamefont {M.}~\bibnamefont
  {Yamaguchi}},\ }\href {\doibase 10.1016/j.physletb.2006.05.078} {\bibfield
  {journal} {\bibinfo  {journal} {Phys. Lett.}\ }\textbf {\bibinfo {volume}
  {B638}},\ \bibinfo {pages} {389} (\bibinfo {year} {2006})},\ \Eprint
  {http://arxiv.org/abs/hep-ph/0602081}{arXiv:hep-ph/0602081
  [hep-ph]}\BibitemShut {NoStop}%
\bibitem [{\citenamefont {Fukugita}\ and\ \citenamefont
  {Yanagida}(1986)}]{Fukugita:1986hr}%
  \BibitemOpen
  \bibfield  {author} {\bibinfo {author} {\bibfnamefont {M.}~\bibnamefont
  {Fukugita}}\ and\ \bibinfo {author} {\bibfnamefont {T.}~\bibnamefont
  {Yanagida}},\ }\href {\doibase 10.1016/0370-2693(86)91126-3} {\bibfield
  {journal} {\bibinfo  {journal} {Phys. Lett.}\ }\textbf {\bibinfo {volume}
  {B174}},\ \bibinfo {pages} {45} (\bibinfo {year} {1986})}\BibitemShut
  {NoStop}%
\bibitem [{\citenamefont {Takahashi}\ \emph {et~al.}(2008)\citenamefont
  {Takahashi}, \citenamefont {Yanagida},\ and\ \citenamefont
  {Yonekura}}]{Takahashi:2008mu}%
  \BibitemOpen
  \bibfield  {author} {\bibinfo {author} {\bibfnamefont {F.}~\bibnamefont
  {Takahashi}}, \bibinfo {author} {\bibfnamefont {T.~T.}\ \bibnamefont
  {Yanagida}}, \ and\ \bibinfo {author} {\bibfnamefont {K.}~\bibnamefont
  {Yonekura}},\ }\href {\doibase 10.1016/j.physletb.2008.05.022} {\bibfield
  {journal} {\bibinfo  {journal} {Phys. Lett.}\ }\textbf {\bibinfo {volume}
  {B664}},\ \bibinfo {pages} {194} (\bibinfo {year} {2008})},\ \Eprint
  {http://arxiv.org/abs/0802.4335}{arXiv:0802.4335 [hep-ph]}\BibitemShut
  {NoStop}%
\bibitem [{\citenamefont {Abbott}\ \emph {et~al.}(2018)\citenamefont {Abbott}
  \emph {et~al.}}]{Abbott:2017xzg}%
  \BibitemOpen
  \bibfield  {author} {\bibinfo {author} {\bibfnamefont {B.~P.}\ \bibnamefont
  {Abbott}} \emph {et~al.} (\bibinfo {collaboration} {Virgo, LIGO
  Scientific}),\ }\href {\doibase 10.1103/PhysRevLett.120.091101} {\bibfield
  {journal} {\bibinfo  {journal} {Phys. Rev. Lett.}\ }\textbf {\bibinfo
  {volume} {120}},\ \bibinfo {pages} {091101} (\bibinfo {year} {2018})},\
  \Eprint {http://arxiv.org/abs/1710.05837}{arXiv:1710.05837
  [gr-qc]}\BibitemShut {NoStop}%
\bibitem [{\citenamefont {Hild}\ \emph {et~al.}(2008)\citenamefont {Hild},
  \citenamefont {Chelkowski},\ and\ \citenamefont {Freise}}]{Hild:2008ng}%
  \BibitemOpen
  \bibfield  {author} {\bibinfo {author} {\bibfnamefont {S.}~\bibnamefont
  {Hild}}, \bibinfo {author} {\bibfnamefont {S.}~\bibnamefont {Chelkowski}}, \
  and\ \bibinfo {author} {\bibfnamefont {A.}~\bibnamefont {Freise}},\
  }\href@noop {} {\  (\bibinfo {year} {2008})},\ \Eprint
  {http://arxiv.org/abs/0810.0604}{arXiv:0810.0604 [gr-qc]}\BibitemShut
  {NoStop}%
\bibitem [{\citenamefont {Sabin}\ \emph {et~al.}(2016)\citenamefont {Sabin},
  \citenamefont {Kohlrus}, \citenamefont {Bruschi},\ and\ \citenamefont
  {Fuentes}}]{Sabin:2015mha}%
  \BibitemOpen
  \bibfield  {author} {\bibinfo {author} {\bibfnamefont {C.}~\bibnamefont
  {Sabin}}, \bibinfo {author} {\bibfnamefont {J.}~\bibnamefont {Kohlrus}},
  \bibinfo {author} {\bibfnamefont {D.~E.}\ \bibnamefont {Bruschi}}, \ and\
  \bibinfo {author} {\bibfnamefont {I.}~\bibnamefont {Fuentes}},\ }\href
  {\doibase 10.1140/epjqt/s40507-016-0046-4} {\bibfield  {journal} {\bibinfo
  {journal} {EPJ Quant. Technol.}\ }\textbf {\bibinfo {volume} {3}},\ \bibinfo
  {pages} {8} (\bibinfo {year} {2016})},\ \Eprint
  {http://arxiv.org/abs/1505.01302}{arXiv:1505.01302 [quant-ph]}\BibitemShut
  {NoStop}%
\bibitem [{\citenamefont {Babu}\ \emph {et~al.}(2016)\citenamefont {Babu},
  \citenamefont {Schmitz},\ and\ \citenamefont {Yanagida}}]{Babu:2015xba}%
  \BibitemOpen
  \bibfield  {author} {\bibinfo {author} {\bibfnamefont {K.~S.}\ \bibnamefont
  {Babu}}, \bibinfo {author} {\bibfnamefont {K.}~\bibnamefont {Schmitz}}, \
  and\ \bibinfo {author} {\bibfnamefont {T.~T.}\ \bibnamefont {Yanagida}},\
  }\href {\doibase 10.1016/j.nuclphysb.2016.01.023} {\bibfield  {journal}
  {\bibinfo  {journal} {Nucl. Phys.}\ }\textbf {\bibinfo {volume} {B905}},\
  \bibinfo {pages} {73} (\bibinfo {year} {2016})},\ \Eprint
  {http://arxiv.org/abs/1507.04467}{arXiv:1507.04467 [hep-ph]}\BibitemShut
  {NoStop}%
\bibitem [{\citenamefont {Bjorkeroth}\ \emph {et~al.}(2017)\citenamefont
  {Bjorkeroth}, \citenamefont {King}, \citenamefont {Schmitz},\ and\
  \citenamefont {Yanagida}}]{Bjorkeroth:2016qsk}%
  \BibitemOpen
  \bibfield  {author} {\bibinfo {author} {\bibfnamefont {F.}~\bibnamefont
  {Bjorkeroth}}, \bibinfo {author} {\bibfnamefont {S.~F.}\ \bibnamefont
  {King}}, \bibinfo {author} {\bibfnamefont {K.}~\bibnamefont {Schmitz}}, \
  and\ \bibinfo {author} {\bibfnamefont {T.~T.}\ \bibnamefont {Yanagida}},\
  }\href {\doibase 10.1016/j.nuclphysb.2017.01.017} {\bibfield  {journal}
  {\bibinfo  {journal} {Nucl. Phys.}\ }\textbf {\bibinfo {volume} {B916}},\
  \bibinfo {pages} {688} (\bibinfo {year} {2017})},\ \Eprint
  {http://arxiv.org/abs/1608.04911}{arXiv:1608.04911 [hep-ph]}\BibitemShut
  {NoStop}%
\bibitem [{\citenamefont {Covi}\ and\ \citenamefont
  {Roulet}(1997)}]{Covi:1996fm}%
  \BibitemOpen
  \bibfield  {author} {\bibinfo {author} {\bibfnamefont {L.}~\bibnamefont
  {Covi}}\ and\ \bibinfo {author} {\bibfnamefont {E.}~\bibnamefont {Roulet}},\
  }\href {\doibase 10.1016/S0370-2693(97)00287-6} {\bibfield  {journal}
  {\bibinfo  {journal} {Phys. Lett.}\ }\textbf {\bibinfo {volume} {B399}},\
  \bibinfo {pages} {113} (\bibinfo {year} {1997})},\ \Eprint
  {http://arxiv.org/abs/hep-ph/9611425}{arXiv:hep-ph/9611425
  [hep-ph]}\BibitemShut {NoStop}%
\bibitem [{\citenamefont {Pilaftsis}(1997)}]{Pilaftsis:1997dr}%
  \BibitemOpen
  \bibfield  {author} {\bibinfo {author} {\bibfnamefont {A.}~\bibnamefont
  {Pilaftsis}},\ }\href {\doibase 10.1016/S0550-3213(97)00469-0} {\bibfield
  {journal} {\bibinfo  {journal} {Nucl. Phys.}\ }\textbf {\bibinfo {volume}
  {B504}},\ \bibinfo {pages} {61} (\bibinfo {year} {1997})},\ \Eprint
  {http://arxiv.org/abs/hep-ph/9702393}{arXiv:hep-ph/9702393
  [hep-ph]}\BibitemShut {NoStop}%
\bibitem [{\citenamefont {Buchmuller}\ and\ \citenamefont
  {Plumacher}(1998)}]{Buchmuller:1997yu}%
  \BibitemOpen
  \bibfield  {author} {\bibinfo {author} {\bibfnamefont {W.}~\bibnamefont
  {Buchmuller}}\ and\ \bibinfo {author} {\bibfnamefont {M.}~\bibnamefont
  {Plumacher}},\ }\href {\doibase 10.1016/S0370-2693(97)01548-7} {\bibfield
  {journal} {\bibinfo  {journal} {Phys. Lett.}\ }\textbf {\bibinfo {volume}
  {B431}},\ \bibinfo {pages} {354} (\bibinfo {year} {1998})},\ \Eprint
  {http://arxiv.org/abs/hep-ph/9710460}{arXiv:hep-ph/9710460
  [hep-ph]}\BibitemShut {NoStop}%
\bibitem [{\citenamefont {Pilaftsis}\ and\ \citenamefont
  {Underwood}(2004)}]{Pilaftsis:2003gt}%
  \BibitemOpen
  \bibfield  {author} {\bibinfo {author} {\bibfnamefont {A.}~\bibnamefont
  {Pilaftsis}}\ and\ \bibinfo {author} {\bibfnamefont {T.~E.~J.}\ \bibnamefont
  {Underwood}},\ }\href {\doibase 10.1016/j.nuclphysb.2004.05.029} {\bibfield
  {journal} {\bibinfo  {journal} {Nucl. Phys.}\ }\textbf {\bibinfo {volume}
  {B692}},\ \bibinfo {pages} {303} (\bibinfo {year} {2004})},\ \Eprint
  {http://arxiv.org/abs/hep-ph/0309342}{arXiv:hep-ph/0309342
  [hep-ph]}\BibitemShut {NoStop}%
\bibitem [{\citenamefont {Pilaftsis}\ and\ \citenamefont
  {Underwood}(2005)}]{Pilaftsis:2005rv}%
  \BibitemOpen
  \bibfield  {author} {\bibinfo {author} {\bibfnamefont {A.}~\bibnamefont
  {Pilaftsis}}\ and\ \bibinfo {author} {\bibfnamefont {T.~E.~J.}\ \bibnamefont
  {Underwood}},\ }\href {\doibase 10.1103/PhysRevD.72.113001} {\bibfield
  {journal} {\bibinfo  {journal} {Phys. Rev.}\ }\textbf {\bibinfo {volume}
  {D72}},\ \bibinfo {pages} {113001} (\bibinfo {year} {2005})},\ \Eprint
  {http://arxiv.org/abs/hep-ph/0506107}{arXiv:hep-ph/0506107
  [hep-ph]}\BibitemShut {NoStop}%
\bibitem [{\citenamefont {Anisimov}\ \emph {et~al.}(2006)\citenamefont
  {Anisimov}, \citenamefont {Broncano},\ and\ \citenamefont
  {Plumacher}}]{Anisimov:2005hr}%
  \BibitemOpen
  \bibfield  {author} {\bibinfo {author} {\bibfnamefont {A.}~\bibnamefont
  {Anisimov}}, \bibinfo {author} {\bibfnamefont {A.}~\bibnamefont {Broncano}},
  \ and\ \bibinfo {author} {\bibfnamefont {M.}~\bibnamefont {Plumacher}},\
  }\href {\doibase 10.1016/j.nuclphysb.2006.01.003} {\bibfield  {journal}
  {\bibinfo  {journal} {Nucl. Phys.}\ }\textbf {\bibinfo {volume} {B737}},\
  \bibinfo {pages} {176} (\bibinfo {year} {2006})},\ \Eprint
  {http://arxiv.org/abs/hep-ph/0511248}{arXiv:hep-ph/0511248
  [hep-ph]}\BibitemShut {NoStop}%
\bibitem [{\citenamefont {Garny}\ \emph {et~al.}(2013)\citenamefont {Garny},
  \citenamefont {Kartavtsev},\ and\ \citenamefont {Hohenegger}}]{Garny:2011hg}%
  \BibitemOpen
  \bibfield  {author} {\bibinfo {author} {\bibfnamefont {M.}~\bibnamefont
  {Garny}}, \bibinfo {author} {\bibfnamefont {A.}~\bibnamefont {Kartavtsev}}, \
  and\ \bibinfo {author} {\bibfnamefont {A.}~\bibnamefont {Hohenegger}},\
  }\href {\doibase 10.1016/j.aop.2012.10.007} {\bibfield  {journal} {\bibinfo
  {journal} {Annals Phys.}\ }\textbf {\bibinfo {volume} {328}},\ \bibinfo
  {pages} {26} (\bibinfo {year} {2013})},\ \Eprint
  {http://arxiv.org/abs/1112.6428}{arXiv:1112.6428 [hep-ph]}\BibitemShut
  {NoStop}%
\bibitem [{\citenamefont {Nakayama}\ \emph {et~al.}(2016)\citenamefont
  {Nakayama}, \citenamefont {Takahashi},\ and\ \citenamefont
  {Yanagida}}]{Nakayama:2016gvg}%
  \BibitemOpen
  \bibfield  {author} {\bibinfo {author} {\bibfnamefont {K.}~\bibnamefont
  {Nakayama}}, \bibinfo {author} {\bibfnamefont {F.}~\bibnamefont {Takahashi}},
  \ and\ \bibinfo {author} {\bibfnamefont {T.~T.}\ \bibnamefont {Yanagida}},\
  }\href {\doibase 10.1016/j.physletb.2016.03.051} {\bibfield  {journal}
  {\bibinfo  {journal} {Phys. Lett.}\ }\textbf {\bibinfo {volume} {B757}},\
  \bibinfo {pages} {32} (\bibinfo {year} {2016})},\ \Eprint
  {http://arxiv.org/abs/1601.00192}{arXiv:1601.00192 [hep-ph]}\BibitemShut
  {NoStop}%
\bibitem [{\citenamefont {Yanagida}(1979)}]{Yanagida:1979as}%
  \BibitemOpen
  \bibfield  {author} {\bibinfo {author} {\bibfnamefont {T.}~\bibnamefont
  {Yanagida}},\ }\bibfield  {booktitle} {\emph {\bibinfo {booktitle}
  {{Proceedings: Workshop on the Unified Theories and the Baryon Number in the
  Universe: Tsukuba, Japan, February 13-14, 1979}}},\ }\href@noop {} {\bibfield
   {journal} {\bibinfo  {journal} {Conf. Proc.}\ }\textbf {\bibinfo {volume}
  {C7902131}},\ \bibinfo {pages} {95} (\bibinfo {year} {1979})}\BibitemShut
  {NoStop}%
\bibitem [{\citenamefont {Gell-Mann}\ \emph {et~al.}(1979)\citenamefont
  {Gell-Mann}, \citenamefont {Ramond},\ and\ \citenamefont
  {Slansky}}]{GellMann:1980vs}%
  \BibitemOpen
  \bibfield  {author} {\bibinfo {author} {\bibfnamefont {M.}~\bibnamefont
  {Gell-Mann}}, \bibinfo {author} {\bibfnamefont {P.}~\bibnamefont {Ramond}}, \
  and\ \bibinfo {author} {\bibfnamefont {R.}~\bibnamefont {Slansky}},\
  }\bibfield  {booktitle} {\emph {\bibinfo {booktitle} {{Supergravity Workshop
  Stony Brook, New York, September 27-28, 1979}}},\ }\href@noop {} {\bibfield
  {journal} {\bibinfo  {journal} {Conf. Proc.}\ }\textbf {\bibinfo {volume}
  {C790927}},\ \bibinfo {pages} {315} (\bibinfo {year} {1979})},\ \Eprint
  {http://arxiv.org/abs/1306.4669}{arXiv:1306.4669 [hep-th]}\BibitemShut
  {NoStop}%
\bibitem [{\citenamefont {Minkowski}(1977)}]{Minkowski:1977sc}%
  \BibitemOpen
  \bibfield  {author} {\bibinfo {author} {\bibfnamefont {P.}~\bibnamefont
  {Minkowski}},\ }\href {\doibase 10.1016/0370-2693(77)90435-X} {\bibfield
  {journal} {\bibinfo  {journal} {Phys. Lett.}\ }\textbf {\bibinfo {volume}
  {67B}},\ \bibinfo {pages} {421} (\bibinfo {year} {1977})}\BibitemShut
  {NoStop}%
\bibitem [{\citenamefont {Shifman}\ and\ \citenamefont
  {Vainshtein}(1988)}]{Shifman:1987ia}%
  \BibitemOpen
  \bibfield  {author} {\bibinfo {author} {\bibfnamefont {M.~A.}\ \bibnamefont
  {Shifman}}\ and\ \bibinfo {author} {\bibfnamefont {A.~I.}\ \bibnamefont
  {Vainshtein}},\ }\href {\doibase 10.1016/0550-3213(88)90680-3} {\bibfield
  {journal} {\bibinfo  {journal} {Nucl. Phys.}\ }\textbf {\bibinfo {volume}
  {B296}},\ \bibinfo {pages} {445} (\bibinfo {year} {1988})},\ \bibinfo {note}
  {[Sov. Phys. JETP66,1100(1987)]}\BibitemShut {NoStop}%
\bibitem [{\citenamefont {Buchmuller}\ \emph {et~al.}(2002)\citenamefont
  {Buchmuller}, \citenamefont {Di~Bari},\ and\ \citenamefont
  {Plumacher}}]{Buchmuller:2002jk}%
  \BibitemOpen
  \bibfield  {author} {\bibinfo {author} {\bibfnamefont {W.}~\bibnamefont
  {Buchmuller}}, \bibinfo {author} {\bibfnamefont {P.}~\bibnamefont {Di~Bari}},
  \ and\ \bibinfo {author} {\bibfnamefont {M.}~\bibnamefont {Plumacher}},\
  }\href {\doibase 10.1016/S0370-2693(02)02758-2} {\bibfield  {journal}
  {\bibinfo  {journal} {Phys. Lett.}\ }\textbf {\bibinfo {volume} {B547}},\
  \bibinfo {pages} {128} (\bibinfo {year} {2002})},\ \Eprint
  {http://arxiv.org/abs/hep-ph/0209301}{arXiv:hep-ph/0209301
  [hep-ph]}\BibitemShut {NoStop}%
\bibitem [{\citenamefont {Takayama}\ and\ \citenamefont
  {Yamaguchi}(2000)}]{Takayama:2000uz}%
  \BibitemOpen
  \bibfield  {author} {\bibinfo {author} {\bibfnamefont {F.}~\bibnamefont
  {Takayama}}\ and\ \bibinfo {author} {\bibfnamefont {M.}~\bibnamefont
  {Yamaguchi}},\ }\href {\doibase 10.1016/S0370-2693(00)00726-7} {\bibfield
  {journal} {\bibinfo  {journal} {Phys. Lett.}\ }\textbf {\bibinfo {volume}
  {B485}},\ \bibinfo {pages} {388} (\bibinfo {year} {2000})},\ \Eprint
  {http://arxiv.org/abs/hep-ph/0005214}{arXiv:hep-ph/0005214
  [hep-ph]}\BibitemShut {NoStop}%
\bibitem [{\citenamefont {Buchmuller}\ \emph {et~al.}(2007)\citenamefont
  {Buchmuller}, \citenamefont {Covi}, \citenamefont {Hamaguchi}, \citenamefont
  {Ibarra},\ and\ \citenamefont {Yanagida}}]{Buchmuller:2007ui}%
  \BibitemOpen
  \bibfield  {author} {\bibinfo {author} {\bibfnamefont {W.}~\bibnamefont
  {Buchmuller}}, \bibinfo {author} {\bibfnamefont {L.}~\bibnamefont {Covi}},
  \bibinfo {author} {\bibfnamefont {K.}~\bibnamefont {Hamaguchi}}, \bibinfo
  {author} {\bibfnamefont {A.}~\bibnamefont {Ibarra}}, \ and\ \bibinfo {author}
  {\bibfnamefont {T.}~\bibnamefont {Yanagida}},\ }\href {\doibase
  10.1088/1126-6708/2007/03/037} {\bibfield  {journal} {\bibinfo  {journal}
  {JHEP}\ }\textbf {\bibinfo {volume} {03}},\ \bibinfo {pages} {037} (\bibinfo
  {year} {2007})},\ \Eprint
  {http://arxiv.org/abs/hep-ph/0702184}{arXiv:hep-ph/0702184
  [HEP-PH]}\BibitemShut {NoStop}%
\bibitem [{\citenamefont {Lee}\ \emph {et~al.}(2011)\citenamefont {Lee},
  \citenamefont {Raby}, \citenamefont {Ratz}, \citenamefont {Ross},
  \citenamefont {Schieren}, \citenamefont {Schmidt-Hoberg},\ and\ \citenamefont
  {Vaudrevange}}]{Lee:2010gv}%
  \BibitemOpen
  \bibfield  {author} {\bibinfo {author} {\bibfnamefont {H.~M.}\ \bibnamefont
  {Lee}}, \bibinfo {author} {\bibfnamefont {S.}~\bibnamefont {Raby}}, \bibinfo
  {author} {\bibfnamefont {M.}~\bibnamefont {Ratz}}, \bibinfo {author}
  {\bibfnamefont {G.~G.}\ \bibnamefont {Ross}}, \bibinfo {author}
  {\bibfnamefont {R.}~\bibnamefont {Schieren}}, \bibinfo {author}
  {\bibfnamefont {K.}~\bibnamefont {Schmidt-Hoberg}}, \ and\ \bibinfo {author}
  {\bibfnamefont {P.~K.~S.}\ \bibnamefont {Vaudrevange}},\ }\href {\doibase
  10.1016/j.physletb.2010.10.038} {\bibfield  {journal} {\bibinfo  {journal}
  {Phys. Lett.}\ }\textbf {\bibinfo {volume} {B694}},\ \bibinfo {pages} {491}
  (\bibinfo {year} {2011})},\ \Eprint
  {http://arxiv.org/abs/1009.0905}{arXiv:1009.0905 [hep-ph]}\BibitemShut
  {NoStop}%
\bibitem [{\citenamefont {Witten}(1982)}]{Witten:1982df}%
  \BibitemOpen
  \bibfield  {author} {\bibinfo {author} {\bibfnamefont {E.}~\bibnamefont
  {Witten}},\ }\href {\doibase 10.1016/0550-3213(82)90071-2} {\bibfield
  {journal} {\bibinfo  {journal} {Nucl. Phys.}\ }\textbf {\bibinfo {volume}
  {B202}},\ \bibinfo {pages} {253} (\bibinfo {year} {1982})}\BibitemShut
  {NoStop}%
\bibitem [{\citenamefont {Zeldovich}\ \emph {et~al.}(1974)\citenamefont
  {Zeldovich}, \citenamefont {Kobzarev},\ and\ \citenamefont
  {Okun}}]{Zeldovich:1974uw}%
  \BibitemOpen
  \bibfield  {author} {\bibinfo {author} {\bibfnamefont {{\relax Ya}.~B.}\
  \bibnamefont {Zeldovich}}, \bibinfo {author} {\bibfnamefont {I.~{\relax
  Yu}.}\ \bibnamefont {Kobzarev}}, \ and\ \bibinfo {author} {\bibfnamefont
  {L.~B.}\ \bibnamefont {Okun}},\ }\href@noop {} {\bibfield  {journal}
  {\bibinfo  {journal} {Zh. Eksp. Teor. Fiz.}\ }\textbf {\bibinfo {volume}
  {67}},\ \bibinfo {pages} {3} (\bibinfo {year} {1974})},\ \bibinfo {note}
  {[Sov. Phys. JETP40,1(1974)]}\BibitemShut {NoStop}%
\bibitem [{\citenamefont {Kibble}(1976)}]{Kibble:1976sj}%
  \BibitemOpen
  \bibfield  {author} {\bibinfo {author} {\bibfnamefont {T.~W.~B.}\
  \bibnamefont {Kibble}},\ }\href {\doibase 10.1088/0305-4470/9/8/029}
  {\bibfield  {journal} {\bibinfo  {journal} {J. Phys.}\ }\textbf {\bibinfo
  {volume} {A9}},\ \bibinfo {pages} {1387} (\bibinfo {year}
  {1976})}\BibitemShut {NoStop}%
\bibitem [{\citenamefont {Vilenkin}(1981)}]{Vilenkin:1981zs}%
  \BibitemOpen
  \bibfield  {author} {\bibinfo {author} {\bibfnamefont {A.}~\bibnamefont
  {Vilenkin}},\ }\href {\doibase 10.1103/PhysRevD.23.852} {\bibfield  {journal}
  {\bibinfo  {journal} {Phys. Rev.}\ }\textbf {\bibinfo {volume} {D23}},\
  \bibinfo {pages} {852} (\bibinfo {year} {1981})}\BibitemShut {NoStop}%
\bibitem [{\citenamefont {Dvali}\ and\ \citenamefont
  {Shifman}(1997)}]{Dvali:1996xe}%
  \BibitemOpen
  \bibfield  {author} {\bibinfo {author} {\bibfnamefont {G.~R.}\ \bibnamefont
  {Dvali}}\ and\ \bibinfo {author} {\bibfnamefont {M.~A.}\ \bibnamefont
  {Shifman}},\ }\href {\doibase 10.1016/S0370-2693(97)00808-3,
  10.1016/S0370-2693(97)00131-7} {\bibfield  {journal} {\bibinfo  {journal}
  {Phys. Lett.}\ }\textbf {\bibinfo {volume} {B396}},\ \bibinfo {pages} {64}
  (\bibinfo {year} {1997})},\ \bibinfo {note} {[Erratum: Phys.
  Lett.B407,452(1997)]},\ \Eprint
  {http://arxiv.org/abs/hep-th/9612128}{arXiv:hep-th/9612128
  [hep-th]}\BibitemShut {NoStop}%
\bibitem [{\citenamefont {Matsuda}(1998)}]{Matsuda:1998ms}%
  \BibitemOpen
  \bibfield  {author} {\bibinfo {author} {\bibfnamefont {T.}~\bibnamefont
  {Matsuda}},\ }\href {\doibase 10.1016/S0370-2693(98)00861-2} {\bibfield
  {journal} {\bibinfo  {journal} {Phys. Lett.}\ }\textbf {\bibinfo {volume}
  {B436}},\ \bibinfo {pages} {264} (\bibinfo {year} {1998})},\ \Eprint
  {http://arxiv.org/abs/hep-ph/9804409}{arXiv:hep-ph/9804409
  [hep-ph]}\BibitemShut {NoStop}%
\bibitem [{\citenamefont {Giudice}\ \emph {et~al.}(2001)\citenamefont
  {Giudice}, \citenamefont {Kolb},\ and\ \citenamefont
  {Riotto}}]{Giudice:2000ex}%
  \BibitemOpen
  \bibfield  {author} {\bibinfo {author} {\bibfnamefont {G.~F.}\ \bibnamefont
  {Giudice}}, \bibinfo {author} {\bibfnamefont {E.~W.}\ \bibnamefont {Kolb}}, \
  and\ \bibinfo {author} {\bibfnamefont {A.}~\bibnamefont {Riotto}},\ }\href
  {\doibase 10.1103/PhysRevD.64.023508} {\bibfield  {journal} {\bibinfo
  {journal} {Phys. Rev.}\ }\textbf {\bibinfo {volume} {D64}},\ \bibinfo {pages}
  {023508} (\bibinfo {year} {2001})},\ \Eprint
  {http://arxiv.org/abs/hep-ph/0005123}{arXiv:hep-ph/0005123
  [hep-ph]}\BibitemShut {NoStop}%
\bibitem [{\citenamefont {Asaka}\ and\ \citenamefont
  {Kawasaki}(1999)}]{Asaka:1999xd}%
  \BibitemOpen
  \bibfield  {author} {\bibinfo {author} {\bibfnamefont {T.}~\bibnamefont
  {Asaka}}\ and\ \bibinfo {author} {\bibfnamefont {M.}~\bibnamefont
  {Kawasaki}},\ }\href {\doibase 10.1103/PhysRevD.60.123509} {\bibfield
  {journal} {\bibinfo  {journal} {Phys. Rev.}\ }\textbf {\bibinfo {volume}
  {D60}},\ \bibinfo {pages} {123509} (\bibinfo {year} {1999})},\ \Eprint
  {http://arxiv.org/abs/hep-ph/9905467}{arXiv:hep-ph/9905467
  [hep-ph]}\BibitemShut {NoStop}%
\bibitem [{\citenamefont {Hiramatsu}\ \emph {et~al.}(2013)\citenamefont
  {Hiramatsu}, \citenamefont {Kawasaki}, \citenamefont {Saikawa},\ and\
  \citenamefont {Sekiguchi}}]{Hiramatsu:2012sc}%
  \BibitemOpen
  \bibfield  {author} {\bibinfo {author} {\bibfnamefont {T.}~\bibnamefont
  {Hiramatsu}}, \bibinfo {author} {\bibfnamefont {M.}~\bibnamefont {Kawasaki}},
  \bibinfo {author} {\bibfnamefont {K.}~\bibnamefont {Saikawa}}, \ and\
  \bibinfo {author} {\bibfnamefont {T.}~\bibnamefont {Sekiguchi}},\ }\href
  {\doibase 10.1088/1475-7516/2013/01/001} {\bibfield  {journal} {\bibinfo
  {journal} {JCAP}\ }\textbf {\bibinfo {volume} {1301}},\ \bibinfo {pages}
  {001} (\bibinfo {year} {2013})},\ \Eprint
  {http://arxiv.org/abs/1207.3166}{arXiv:1207.3166 [hep-ph]}\BibitemShut
  {NoStop}%
\bibitem [{\citenamefont {Hiramatsu}\ \emph {et~al.}(2014)\citenamefont
  {Hiramatsu}, \citenamefont {Kawasaki},\ and\ \citenamefont
  {Saikawa}}]{Hiramatsu:2013qaa}%
  \BibitemOpen
  \bibfield  {author} {\bibinfo {author} {\bibfnamefont {T.}~\bibnamefont
  {Hiramatsu}}, \bibinfo {author} {\bibfnamefont {M.}~\bibnamefont {Kawasaki}},
  \ and\ \bibinfo {author} {\bibfnamefont {K.}~\bibnamefont {Saikawa}},\ }\href
  {\doibase 10.1088/1475-7516/2014/02/031} {\bibfield  {journal} {\bibinfo
  {journal} {JCAP}\ }\textbf {\bibinfo {volume} {1402}},\ \bibinfo {pages}
  {031} (\bibinfo {year} {2014})},\ \Eprint
  {http://arxiv.org/abs/1309.5001}{arXiv:1309.5001 [astro-ph.CO]}\BibitemShut
  {NoStop}%
\bibitem [{\citenamefont {Dine}\ \emph {et~al.}(2010)\citenamefont {Dine},
  \citenamefont {Takahashi},\ and\ \citenamefont {Yanagida}}]{Dine:2010eb}%
  \BibitemOpen
  \bibfield  {author} {\bibinfo {author} {\bibfnamefont {M.}~\bibnamefont
  {Dine}}, \bibinfo {author} {\bibfnamefont {F.}~\bibnamefont {Takahashi}}, \
  and\ \bibinfo {author} {\bibfnamefont {T.~T.}\ \bibnamefont {Yanagida}},\
  }\href {\doibase 10.1007/JHEP07(2010)003} {\bibfield  {journal} {\bibinfo
  {journal} {JHEP}\ }\textbf {\bibinfo {volume} {07}},\ \bibinfo {pages} {003}
  (\bibinfo {year} {2010})},\ \Eprint
  {http://arxiv.org/abs/1005.3613}{arXiv:1005.3613 [hep-th]}\BibitemShut
  {NoStop}%
\bibitem [{\citenamefont {Harigaya}\ \emph {et~al.}(2015)\citenamefont
  {Harigaya}, \citenamefont {Ibe}, \citenamefont {Schmitz},\ and\ \citenamefont
  {Yanagida}}]{Harigaya:2015yla}%
  \BibitemOpen
  \bibfield  {author} {\bibinfo {author} {\bibfnamefont {K.}~\bibnamefont
  {Harigaya}}, \bibinfo {author} {\bibfnamefont {M.}~\bibnamefont {Ibe}},
  \bibinfo {author} {\bibfnamefont {K.}~\bibnamefont {Schmitz}}, \ and\
  \bibinfo {author} {\bibfnamefont {T.~T.}\ \bibnamefont {Yanagida}},\ }\href
  {\doibase 10.1016/j.physletb.2015.07.073} {\bibfield  {journal} {\bibinfo
  {journal} {Phys. Lett.}\ }\textbf {\bibinfo {volume} {B749}},\ \bibinfo
  {pages} {298} (\bibinfo {year} {2015})},\ \Eprint
  {http://arxiv.org/abs/1506.00426}{arXiv:1506.00426 [hep-ph]}\BibitemShut
  {NoStop}%
\bibitem [{\citenamefont {Kawasaki}\ \emph {et~al.}(2015)\citenamefont
  {Kawasaki}, \citenamefont {Saikawa},\ and\ \citenamefont
  {Sekiguchi}}]{Kawasaki:2014sqa}%
  \BibitemOpen
  \bibfield  {author} {\bibinfo {author} {\bibfnamefont {M.}~\bibnamefont
  {Kawasaki}}, \bibinfo {author} {\bibfnamefont {K.}~\bibnamefont {Saikawa}}, \
  and\ \bibinfo {author} {\bibfnamefont {T.}~\bibnamefont {Sekiguchi}},\ }\href
  {\doibase 10.1103/PhysRevD.91.065014} {\bibfield  {journal} {\bibinfo
  {journal} {Phys. Rev.}\ }\textbf {\bibinfo {volume} {D91}},\ \bibinfo {pages}
  {065014} (\bibinfo {year} {2015})},\ \Eprint
  {http://arxiv.org/abs/1412.0789}{arXiv:1412.0789 [hep-ph]}\BibitemShut
  {NoStop}%
\bibitem [{\citenamefont {Gleiser}\ and\ \citenamefont
  {Roberts}(1998)}]{Gleiser:1998na}%
  \BibitemOpen
  \bibfield  {author} {\bibinfo {author} {\bibfnamefont {M.}~\bibnamefont
  {Gleiser}}\ and\ \bibinfo {author} {\bibfnamefont {R.}~\bibnamefont
  {Roberts}},\ }\href {\doibase 10.1103/PhysRevLett.81.5497} {\bibfield
  {journal} {\bibinfo  {journal} {Phys. Rev. Lett.}\ }\textbf {\bibinfo
  {volume} {81}},\ \bibinfo {pages} {5497} (\bibinfo {year} {1998})},\ \Eprint
  {http://arxiv.org/abs/astro-ph/9807260}{arXiv:astro-ph/9807260
  [astro-ph]}\BibitemShut {NoStop}%
\bibitem [{\citenamefont {Takahashi}\ and\ \citenamefont
  {Yanagida}(2011)}]{Takahashi:2011as}%
  \BibitemOpen
  \bibfield  {author} {\bibinfo {author} {\bibfnamefont {F.}~\bibnamefont
  {Takahashi}}\ and\ \bibinfo {author} {\bibfnamefont {T.~T.}\ \bibnamefont
  {Yanagida}},\ }\href {\doibase 10.1016/j.physletb.2011.03.032} {\bibfield
  {journal} {\bibinfo  {journal} {Phys. Lett.}\ }\textbf {\bibinfo {volume}
  {B698}},\ \bibinfo {pages} {408} (\bibinfo {year} {2011})},\ \Eprint
  {http://arxiv.org/abs/1101.0867}{arXiv:1101.0867 [hep-ph]}\BibitemShut
  {NoStop}%
\end{thebibliography}

%

}

\end{document}